\documentclass[traditabstract]{aa}  
\usepackage{graphicx}
\usepackage{txfonts}
\usepackage{natbib}
\bibpunct{(}{)}{;}{a}{}{,}
\begin{document}
   \title{The HARPS search for southern extrasolar planets.\thanks{Based on observations made with the HARPS instrument on the ESO 
   3.6-m telescope at  the La Silla Observatory (Chile) under the GTO programme ID 072.C-0488.}
   \fnmsep\thanks{Radial-velocity tables are only available in electronic form at the CDS via anonymous ftp to cdsarc.u-strasbg.fr 
   (130.79.128.5) or via http://cdsweb.u-strasbg.fr/cgi-bin/qcat?J/A+A/}
         }

   \subtitle{XXIII. 8 planetary companions to low-activity solar-type stars.}

   \author{
              D.~Naef\inst{\ref{inst1}}
              \and
              M.~Mayor\inst{\ref{inst1}}
              \and
              G.~Lo~Curto\inst{\ref{inst2}}
              \and
              F.~Bouchy\inst{\ref{inst3},\ref{inst4}}
              \and
              C.~Lovis\inst{\ref{inst1}}
              \and
              C.~Moutou\inst{\ref{inst5}}
              \and
              W.~Benz\inst{\ref{inst6}}
              \and
              F.~Pepe\inst{\ref{inst1}}
              \and
              D.~Queloz\inst{\ref{inst1}}
              \and
              N.C.~Santos\inst{\ref{inst7}}
              \and
              D.~S\'{e}gransan\inst{\ref{inst1}}
              \and
              S.~Udry\inst{\ref{inst1}}
              \and
	      X.~Bonfils\inst{\ref{inst1},\ref{inst8}}
	      \and
	      X.~Delfosse\inst{\ref{inst8}}
	      \and
	      T.~Forveille\inst{\ref{inst8}}
	      \and
              G.~H\'{e}brard\inst{\ref{inst3}}
	      \and
	      C.~Mordasini\inst{\ref{inst6},\ref{inst9}}
              \and
	      C.~Perrier\inst{\ref{inst8}}
	      \and
	      I.~Boisse\inst{\ref{inst3}}
	      \and
	      D.~Sosnowska\inst{\ref{inst1}}
          }

   \institute{   Observatoire Astronomique de l'Universit\'{e} de Gen\`{e}ve, 51 Ch. des Maillettes, CH--1290 Sauverny,
                 Switzerland. \email{dominique.naef@unige.ch}\label{inst1}
                 \and
                 European Southern Observatory, Karl Schwarzschild Strasse 2, D--85748 Garching bei M\"{u}nchen, Germany.\label{inst2}
                 \and
		 Institut d'Astrophysique de Paris, UMR7095 CNRS, Universit\'e Pierre \& Marie Curie, 98bis boulevard Arago, F--75014 Paris, France.\label{inst3}
                 \and
		 Observatoire de Haute-Provence, F--04870 St-Michel L'Observatoire, France.\label{inst4}
		 \and 
                 Laboratoire d'Astrophysique de Marseille, OAMP, Universit\'e Aix-Marseille \&                
                 CNRS, 38 rue Fr\'ed\'eric Jolliot-Curie, F--13388 Marseille cedex 13, France.\label{inst5}
		 \and
                 Physikalisches Institut Universit\"{a}t Bern, Sidlerstrasse 5, CH--3012 Bern, 
                 Switzerland.\label{inst6}
                 \and
                 Centro de Astrof\'{i}sica, Universidade do Porto, Rua das Estrelas, P--4150-762 
                 Porto, Portugal.\label{inst7}
		 \and
		 Laboratoire d'Astrophysique de Grenoble, UMR5571 CNRS, Universit\'e Joseph Fourrier, BP\,53,
		 F--38041 Grenoble cedex 9, France.\label{inst8}
		 \and
		 Max-Planck-Institut f\"{u}r Astronomie, K\"{o}nigstuhl 17, D--69117 Heidelberg, Germany.\label{inst9}
             }

   \date{Received 6 November 2009 / Accepted 10 August 2010}

\abstract{In this paper, we present our {\footnotesize HARPS} radial-velocity data for eight low-activity solar-type stars belonging to 
the {\footnotesize HARPS} volume-limited sample: \object{{\footnotesize HD}\,6718}, \object{{\footnotesize HD}\,8535}, 
\object{{\footnotesize HD}\,28254}, {\footnotesize HD}\,290327, \object{{\footnotesize HD}\,43197}, \object{{\footnotesize HD}\,44219}, 
\object{{\footnotesize HD}\,148156}, and \object{{\footnotesize HD}\,156411}. Keplerian fits to these data reveal the presence of 
low-mass companions around these targets. With minimum masses ranging from 0.58 to 2.54\,M$_{\rm Jup}$, these companions are in the 
planetary mass domain. The orbital periods of these planets range from slightly less than one to almost seven years. The eight orbits 
presented in this paper exhibit a wide variety of eccentricities: from 0.08 to above 0.8. 
}


\maketitle
\titlerunning{The HARPS search for southern extra-solar planets}
\authorrunning{Naef et al.}
%

\section{Introduction}\label{intro}

In this paper, we present new detections from the {\footnotesize HARPS} planet search programme. It has been ongoing since 2003. Its 
detections are based on radial velocities obtained with the {\it High Accuracy Radial velocity Planet Searcher} 
\cite[{\footnotesize HARPS,}][]{Mayor2003}.  {\footnotesize HARPS} is a high-resolution echelle spectrograph mounted on the 
{\footnotesize ESO}--3.6-m telescope in La Silla (Chile). The instrument is installed in a vacuum vessel, and it is thermally controlled. 
These characteristics result in a very high stability of the intrumental radial-velocity zero point (measured drifts smaller than 
1\,m\,s$^{\rm -1}$ over one night). {\footnotesize HARPS} possesses two optical fibres. The first one collects the target light, whereas 
the second one can be used for recording a reference Thorium-Argon spectrum. The second spectrum has been used for measuring the residual 
drift of the instrument since the time of the latest wavelength calibration, allowing us to apply a correction to the velocities 
\citep{Baranne96}. The second fibre can also be used for measuring the spectrum of the sky background. With more than 70 detections 
obtained so far, including the lightest planet \citep[\object{{\footnotesize Gl}\,581}\,e, ][]{Mayor2009}, the {\footnotesize HARPS} 
contribution to the exoplanet research field is a major one. 

Since its beginning, the {\footnotesize HARPS} planet search programme has been divided into several sub-programmes related to specific 
scientific questions. One of these sub-programmes is the {\sl {\footnotesize HARPS} volume-limited programme} to which about 20\% of the 
{\footnotesize HARPS} Guaranteed Time Observation ({\footnotesize GTO}) programme allocated time was dedicated. The sample of the 
{\footnotesize HARPS} volume-limited programme is an extension up to 57.5\,pc of the {\footnotesize CORALIE} planet search sample 
\citep[volume-limited sample up to 50\,pc, see ][]{Udrysample}. It contains about 850 F8--M0 dwarfs. As in the case of the 
{\footnotesize CORALIE} sample, the maximum distance decreases with the $B-V$ colour index for targets with late spectral types (later 
than K0). Our target list will be published in another paper (Lo~Curto et al., in prep.).

Targets in this sample are observed at  lower signal-to-noise ratios (typically 40-50) than the ones in the high-precision programme and 
without using the simultaneous calibration. The typical photon-noise error obtained in this case is about 1-2\,m\,s$^{\rm -1}$ depending 
on the target spectral type and projected rotational velocity. This strategy, optimized for the follow-up of a large stellar sample, is 
a good compromise between precision and observing efficiency.

\begin{table*}[ht!]
\caption{
\label{tabstars}
Observed and inferred stellar characteristics of the host stars.
}
\begin{tabular}{ll|r@{\,$\pm$\,}lr@{\,$\pm$\,}lr@{\,$\pm$\,}lr@{\,$\pm$\,}l}
\hline\hline
\noalign{\vspace{0.05cm}}
{\bf Parameter}           & {\bf Unit}          & \multicolumn{2}{c}{\bf HD\,6718}                 & \multicolumn{2}{c}{\bf HD\,8535}                 & \multicolumn{2}{c}{\bf HD\,28254}                & \multicolumn{2}{c}{\bf HD\,290327}\\
\hline
\noalign{\vspace{0.05cm}}
$Sp.\,Type$               &                     & \multicolumn{2}{c}{G5V}                          & \multicolumn{2}{c}{G0V}                          & \multicolumn{2}{c}{G1IV/V}                       & \multicolumn{2}{c}{G8V}\\[0.1cm]
$m_{\rm V}$               &                     & \multicolumn{2}{c}{8.45}                         & \multicolumn{2}{c}{7.70}                         & \multicolumn{2}{c}{7.69}                         & \multicolumn{2}{c}{8.99}\\[0.1cm]
$B-V$                     &                     & \multicolumn{2}{c}{0.662}                        & \multicolumn{2}{c}{0.553}                        & \multicolumn{2}{c}{0.722}                        & \multicolumn{2}{c}{0.761}\\[0.1cm]
$\pi$                     & (mas)               & 18.23 & 0.76                                     & 19.03 & 0.60                                     & 18.29 & 0.53                                     & 17.65 & 1.57\\[0.1cm]
$d$                       & (pc)                & \multicolumn{2}{c}{54.9$^{\rm +2.4}_{-\rm 2.2}$} & \multicolumn{2}{c}{52.5$^{\rm +1.7}_{-\rm 1.6}$} & \multicolumn{2}{c}{54.7$^{\rm +1.6}_{-\rm 1.5}$} & \multicolumn{2}{c}{56.7$^{\rm +5.5}_{-\rm 4.6}$}\\[0.1cm]
$M_{\rm V}$               &                     & \multicolumn{2}{c}{4.754}                        & \multicolumn{2}{c}{4.097}                        & \multicolumn{2}{c}{4.001}                        & \multicolumn{2}{c}{5.224}\\[0.1cm]
$B.C.$                    &                     & \multicolumn{2}{c}{$-$0.145}                     & \multicolumn{2}{c}{$-$0.028}                     & \multicolumn{2}{c}{$-$0.102}                     & \multicolumn{2}{c}{$-$0.126}\\[0.1cm]
$L$                       & (L$_{\odot}$)       & \multicolumn{2}{c}{1.13}                         & \multicolumn{2}{c}{1.86}                         & \multicolumn{2}{c}{2.18}                         & \multicolumn{2}{c}{0.72}\\[0.1cm]
$T_{\rm eff}$             & (K)                 & 5746    & 19                                     & 6136 & 18                                        & 5664 & 35                                        & 5552    & 21\\[0.1cm]
$\log g$\tablefootmark{a} & (cgs)               & 4.48    & 0.03                                   & 4.46 & 0.06                                      & 4.12 & 0.05                                      & 4.42    & 0.04\\[0.1cm]
$\log g$\tablefootmark{b} & (cgs)               & \multicolumn{2}{c}{4.40}                         & \multicolumn{2}{c}{4.35}                         & \multicolumn{2}{c}{4.13}                         & \multicolumn{2}{c}{4.39}\\[0.1cm]
$[$Fe/H$]$                &                     & $-$0.06 & 0.02                                   & 0.06 & 0.02                                      & 0.36 & 0.03                                      & $-$0.11 & 0.02\\[0.1cm]
$M_*$                     & (M$_{\odot}$)       & \multicolumn{2}{c}{0.96}                         & \multicolumn{2}{c}{1.13}                         & \multicolumn{2}{c}{1.06}                         & \multicolumn{2}{c}{0.90}\\[0.1cm]
$R_*$                     & (R$_{\odot}$)       & 1.02    & 0.03                                   & 1.19 & 0.04                                      & 1.48 & 0.06                                      & 1.00    & 0.01\\[0.1cm]
$\log R^{'}_{HK}$         &                     & \multicolumn{2}{c}{$-$4.97}                      & \multicolumn{2}{c}{$-$4.95}                      & \multicolumn{2}{c}{$-$5.10}                      & \multicolumn{2}{c}{$-$4.96}\\[0.1cm]
$v\sin i$                 & (km\,s$^{\rm -1}$)  & 1.76 & 1.0                                       & 1.41 & 1.0                                       & 2.50 & 1.0                                       & 1.44 & 1.0\\[0.05cm]
\hline
\noalign{\vspace{0.05cm}}
{\bf Parameter}           & {\bf Unit}          & \multicolumn{2}{c}{\bf HD\,43197}                & \multicolumn{2}{c}{\bf HD\,44219}                & \multicolumn{2}{c}{\bf HD\,148156}               & \multicolumn{2}{c}{\bf HD\,156411}\\
\hline
\noalign{\vspace{0.05cm}}
$Sp.\,Type$               &                     & \multicolumn{2}{c}{G8V}                          & \multicolumn{2}{c}{G2V}                          & \multicolumn{2}{c}{F8V}                          & \multicolumn{2}{c}{F8IV/V}\\[0.1cm]
$m_{\rm V}$               &                     & \multicolumn{2}{c}{8.98}                         & \multicolumn{2}{c}{7.69}                         & \multicolumn{2}{c}{7.69}                         & \multicolumn{2}{c}{6.67}\\[0.1cm]
$B-V$                     &                     & \multicolumn{2}{c}{0.817}                        & \multicolumn{2}{c}{0.687}                        & \multicolumn{2}{c}{0.560}                        & \multicolumn{2}{c}{0.614}\\[0.1cm]
$\pi$                     & (mas)               & 17.76 & 1.22                                     & 19.83 & 0.78                                     & 19.38 & 0.75                                     & 18.25 & 0.49\\[0.1cm]
$d$                       & (pc)                & \multicolumn{2}{c}{56.3$^{\rm +4.2}_{-\rm 3.6}$} & \multicolumn{2}{c}{50.4$^{\rm +2.1}_{-\rm 1.9}$} & \multicolumn{2}{c}{51.6$^{\rm +2.1}_{-\rm 1.9}$} & \multicolumn{2}{c}{54.8$^{\rm +1.5}_{-\rm 1.4}$}\\[0.1cm]
$M_{\rm V}$               &                     & \multicolumn{2}{c}{5.227}                        & \multicolumn{2}{c}{4.177}                        & \multicolumn{2}{c}{4.127}                        & \multicolumn{2}{c}{2.976}\\[0.1cm]
$B.C.$                    &                     & \multicolumn{2}{c}{$-$0.137}                     & \multicolumn{2}{c}{$-$0.085}                     & \multicolumn{2}{c}{$-$0.009}                     & \multicolumn{2}{c}{$-$0.060}\\[0.1cm]
$L$                       & (L$_{\odot}$)       & \multicolumn{2}{c}{0.73}                         & \multicolumn{2}{c}{1.82}                         & \multicolumn{2}{c}{1.78}                         & \multicolumn{2}{c}{5.38}\\[0.1cm]
$T_{\rm eff}$             & (K)                 & 5508 & 46                                        & 5752 & 16                                        & 6308 & 28                                        & 5900    & 15\\[0.1cm]
$\log g$\tablefootmark{a} & (cgs)               & 4.31 & 0.08                                      & 4.21 & 0.03                                      & 4.56 & 0.07                                      & 4.07    & 0.04\\[0.1cm]
$\log g$\tablefootmark{b} & (cgs)               & \multicolumn{2}{c}{4.42}                         & \multicolumn{2}{c}{4.20}                         & \multicolumn{2}{c}{4.36}                         & \multicolumn{2}{c}{3.87}\\[0.1cm]
$[$Fe/H$]$                &                     & 0.40 & 0.04                                      & 0.03 & 0.01                                      & 0.29 & 0.02                                      & $-$0.12 & 0.01\\[0.1cm]
$M_*$                     & (M$_{\odot}$)       & \multicolumn{2}{c}{0.96}                         & \multicolumn{2}{c}{1.00}                         & \multicolumn{2}{c}{1.22}                         & \multicolumn{2}{c}{1.25}\\[0.1cm]
$R_*$                     & (R$_{\odot}$)       & 1.00 & 0.01                                      & 1.32 & 0.07                                      & 1.21 & 0.03                                      & 2.16 & 0.09\\[0.1cm]
$\log R^{'}_{HK}$         &                     & \multicolumn{2}{c}{$-$5.06}                      & \multicolumn{2}{c}{$-$5.03}                      & \multicolumn{2}{c}{$-$4.94}                      & \multicolumn{2}{c}{$-$5.05}\\[0.1cm]
$v\sin i$                 & (km\,s$^{\rm -1}$)  & 2.18 & 1.0                                       & 2.22 & 1.0                                       & 5.70 & 1.0                                       & 3.30 & 1.0\\[0.05cm]
\hline
\end{tabular}
\tablefoot{
\tablefoottext{c}{From our spectroscopic {\footnotesize LTE} analysis}
\tablefoottext{d}{Using $M_*$ and $R_*$}}
\end{table*}

Rather short exposure times (less than 5 minutes) are usually enough for reaching the targetted signal-to-noise ratio. An exposure time of 
5 minutes is too short for correctly averaging out intrinsic stellar signals due for example to pulsation or granulation. Our velocities 
can thus be affected by those effects that in some cases can reach a peak-to-peak amplitude of several m\,s$^{\rm -1}$ 
\citep[see e.g. the case of \object{$\mu$\,Arae}, ][]{Santosmuarae, Bouchy2005}. The amplitudes of these signals depend on the target 
spectral types and on the averaging strategy. Expected radial-velocity errors due to intrinsic signals are greater for earlier spectral 
types and for evolved targets.

With such an observing strategy, our programme is mostly sensitive to Saturn or Jupiter-like planets but is also sensitive to lower-mass 
planets around stars with a low intrinsic radial-velocity variability. Also, we generally avoid taking exposures shorter than 1 minute in 
order to reduce the impact of observing overheads. As a consequence, the brightest targets in our sample frequently have photon-noise 
errors significantly lower than the targetted one. For illustrating these facts, we mention the recent detections of a 5.5\,M$_{\oplus}$ 
planet around \object{{\footnotesize HD}\,215497} \citep{Locurto2009}, of the 12.6\,M$_{\oplus}$ planet 
\object{{\footnotesize HD}\,125595}\,b \citep{Segransan2009} and of the 14.4\,M$_{\oplus}$ planet around 
\object{{\footnotesize BD}\,$-$08\,2823}\,b \citep{Hebrard2009}.

The main aim of this observational effort is a better characterization of orbital elements distributions of gaseous planets. With the 
increasing size of the sample of known planets, a quantitative comparison between the observed sample and the predictions of planet 
formation models becomes possible and reliable \citep[see for example][]{Mordasini2009}. With the eight planetary companions presented in 
this paper, the total number of planets detected in the volume-limited sample now reaches 39 
\citep[see][]{Pepe2004, Moutou2005, Locurto2006, Naef2007, Moutou2009, Locurto2009, MordasiniHARPS, Segransan2009, Moutou2010}.

In this paper, we present our {\footnotesize HARPS} radial-velocity data for 8 solar-type stars hosting low-mass companions: 
\object{{\footnotesize HD}\,6718}, \object{{\footnotesize HD}\,8535}, \object{{\footnotesize HD}\,28254}, {\footnotesize HD}\,290327, 
\object{{\footnotesize HD}\,43197}, \object{{\footnotesize HD}\,44219}, \object{{\footnotesize HD}\,148156}, and 
\object{{\footnotesize HD}\,156411}. This paper is organized as follows. We present the main characteristics of the 8 host stars in 
Sect.~\ref{stars}. Section~\ref{rvdata} contains the description of the radial-velocity sets and orbital solutions obtained for these 
targets. Finally, we provide concluding remarks in Sect.~\ref{conclusion}.

\section{Characteristics of the host stars}\label{stars}

We list the main characteristics of the nine host stars in Table~\ref{tabstars}. Apparent V-band magnitudes $m_{\rm V}$ and colour indexes 
$B-V$ are from the {\footnotesize HIPPARCOS} Catalogue \citep{ESA97}. Distances $d$ are based on the improved {\footnotesize HIPPARCOS} 
astrometric parallaxes $\pi$  derived by \citet{vanleeuwen2007}. Absolute magnitudes $M_{\rm V}$ were computed using $m_{\rm V}$ and $d$. 

For each star, we derived the atmospheric parameters $T_{\rm eff}$, $\log g$, and $[$Fe/H$]$ by performing a detailed local 
thermodynamical equilibrium spectroscopic analysis. A high signal-to-noise co-added {\footnotesize HARPS} spectrum was built for each 
target. The set of \ion{Fe}{I}/\ion{Fe}{II} equivalent widths used for the analysis were measured on the co-added spectra with the 
{\footnotesize ARES} code \citep{Sousa2007}. The list of spectral lines for the analysis, as well as more details on the method can be 
found in \citet{Santos2004}.  

Stellar luminosities $L$ were computed using {\footnotesize HIPPARCOS} apparent magnitudes and parallaxes and the bolometric correction 
obtained via the $T_{\rm eff}$-based calibration in \citet{Flower96}. Stellar masses $M_{\rm *}$, and radii $R_{\rm *}$ were derived from 
$T_{\rm eff}$, $[$Fe/H$]$, and $M_{\rm V}$ by interpolating the \citet{Girardi2000} theoretical isochrones. Following 
\citet{Fernandes2004}, a relative uncertainty of $\simeq$ 10\% can be assumed for $M_{\rm *}$. We also computed surface gravities, 
$\log g$, using $M_{\rm *}$ and $R_{\rm *}$. These values are in fair agreement with the spectroscopic ones: a mean difference 
(spectroscopy$-$isochrones) of 0.06\,dex and an {\sl rms} of the differences equal to 0.11\,dex. Spectral types listed in 
Table\ref{tabstars} are based on the stellar parameters  we determined in this paper.  \citet{Jenkins2008} have published iron abundances 
for \object{{\footnotesize HD}\,43197} and \object{{\footnotesize HD}\,148156}. Their values, $[$Fe/H$]$\,=\,0.37\,$\pm$\,0.09 and 
0.22\,$\pm$\,0.15, respectively, are in good agreement with ours. The atmospheric parameters for \object{{\footnotesize HD}\,44219} have 
also been determined by \citet{Robinson2007}. They found $T_{\rm eff}$\,=\,5723\,$\pm$\,82, $\log g$\,=\,4.37\,$\pm$\,0.13 and 
$[$Fe/H$]$\,=\,$-$0.05\,$\pm$\,0.07. These values agree with ours within uncertainties.

Activity indexes $\log R^{'}_{\rm HK}$  were extracted from individual {\footnotesize HARPS} spectra using a method similar to the one 
described in \citet{Santos2000}. Values indicated in Table~\ref{tabstars} are average activity indexes obtained for each target. All the 
targets exhibit a very low level of activity. Our $\log R^{'}_{\rm HK}$ values for \object{{\footnotesize HD}\,43197} and 
\object{{\footnotesize HD}\,148156} are in good agreement with those published in \citet{Jenkins2008}: $-$5.12 and $-$5.08, respectively. 
We did not try to estimate chromospheric ages for our targets because these two quantities are no longer correlated for weakly active 
stars \citep{Pace2004}. However, we can infer for most of these stars (if not all) a safe age constraint based on their low activity 
indexes and rather high luminosities: $age$\,$>$\,3\,Gyr. The projected rotational velocities $v\sin i$ were derived from the widths of 
the {\footnotesize HARPS} cross-correlation functions ({\footnotesize CCF}) following the method presented in \citet{Santos2002}. 
  
Finally, \object{{\footnotesize HD}\,28254} is the bright component of a visual binary. According to the Catalogue of Components of Double 
and Multiple stars \citep{Dommanget2002}, the visual companion, \object{{\footnotesize CCDM}\,J04248-5037\,B}, is located at a separation 
of 4.3\arcsec  and at a position angle of 254 degrees. With an apparent magnitude of  $m_{\rm V}$\,=\,13.8, this object is 6.11\,mag 
fainter (i.e. a flux ratio less than 0.4\%). It was thus clearly outside the detection capabilities of {\footnotesize HIPPARCOS}. Assuming 
the two stars form a bound system and using the distance of \object{{\footnotesize HD}\,28254} (54.7\,pc), we computed the projected 
separation between them: 235\,AU.  We also computed the absolute magnitude of the B component: $M_{\rm V}$\,=\,10.11 corresponding to an 
M0V-M2V star. With this absolute magnitiude and the mass-luminosity relations in \citet{Delfosse2000}, we computed a mass of 
0.48\,M$_\odot$ for this object. Its minimum period is about 1000\,yr. Assuming that this companion is on a circular edge-on orbit, the 
maximal radial-velocity amplitude it would induce on the primary star is of the order of 1.1 km\,s$^{\rm -1}$.

\section{HARPS radial-velocity data and orbital solutions}\label{rvdata}

As explained before, stars belonging to the {\footnotesize HARPS} volume-limited sample are generally observed without simultaneous 
thorium-argon reference. Possible instrumental drifts therefore remain uncorrected. This has only a limited impact on our results because: 
1) the instrument is very stable: on average, the radial-velocity drifts of {\footnotesize HARPS} remain below 0.5\,m\,s$^{\rm -1}$ over 
one night and 2) the targetted photon-noise error for this programme is 1-2\,m\,s$^{\rm -1}$. We stress again the fact that our observing 
strategy is not optimally designed for averaging out stellar intrinsic signals. We list in Table~\ref{rvsol} the median exposure times for 
all the targets. This median is below 3\,minutes in all cases. We thus expect our velocities to be affected at some level by stellar 
intrinsic jitter in particular the velocities of our early type and/or slightly evolved targets, namely \object{{\footnotesize HD}\,8535} 
(G0V), \object{{\footnotesize HD}\,28254} (G1IV/V), \object{{\footnotesize HD}\,148156} (F8V), and \object{{\footnotesize HD}\,156411} 
(F8IV/V).

Our radial velocities can be obtained from the extracted spectra by cross-correlating them with a numerical template. Numerical templates 
are chosen to minimize the mismatch with the target spectral type. Our radial-velocity uncertainties include two well-quantified 
contributions: photon noise and calibration error. An additional error of 0.5\,m\,s$^{\rm -1}$ is added quadratically to account for the 
absence of simultanous calibration and thus the possibility of a small uncorrected radial-velocity drift of the {\footnotesize HARPS} zero 
point. 

All the orbital solutions presented in the following sections were obtained with software that includes a genetic algorithm. In general, 
the only a priori information used by this algorithm is the type of model we want to use (i.e., 1-Keplerian, 2-Keplerian, 1-Keplerian plus 
a linear/quadratic drift, etc.). Once the global parameter space minimum is identified, we use it as the starting point for a standard 
Levenberg-Marquardt minimization out of which the final solution is obtained. In all the cases, we have verified the absence of 
correlation between the post-fit residuals and the measured cross-correlation function bisectors. Finally, uncertainities on the fitted 
parameters are determined using Monte-Carlo simulations.

\begin{table*}[t!]
\caption{
\label{rvsol}
Orbital solutions for the eight host stars. 
}
\begin{tabular}{ll|r@{\,$\pm$\,}lr@{\,$\pm$\,}lr@{\,$\pm$\,}lr@{\,$\pm$\,}l}
\hline\hline
\noalign{\vspace{0.05cm}}
{\bf Parameter}                     & {\bf Unit}                       & \multicolumn{2}{c}{\bf HD\,6718}                   & \multicolumn{2}{c}{\bf HD\,8535}                           & \multicolumn{2}{c}{\bf HD\,28254}                  & \multicolumn{2}{c}{\bf HD\,290327}\\
\hline
\noalign{\vspace{0.05cm}}
$P$                                 & (d)                              & 2496    & 176                                      & 1313   & 28                                                & 1116 & 26                                          & \multicolumn{2}{c}{2443$^{\rm +205}_{-\rm 117}$}\\[0.1cm]
$T$                                 & (JD\tablefootmark{a})            & 357     & 251                                      & 537    & 92                                                & \multicolumn{2}{c}{49$^{\rm +43}_{-\rm 55}$}       & \multicolumn{2}{c}{1326$^{\rm +230}_{-\rm 412}$}\\[0.1cm]
$e$                                 &                                  & \multicolumn{2}{c}{0.10$^{\rm+0.11}_{-\rm 0.04}$}  & \multicolumn{2}{c}{0.15$^{\rm+0.09}_{-\rm 0.05}$}          & \multicolumn{2}{c}{0.81$^{\rm+0.05}_{-\rm 0.02}$}  & \multicolumn{2}{c}{0.08$^{\rm +0.08}_{-\rm 0.03}$}\\[0.1cm]
$\gamma$                            & (km\,s$^{\rm -1}$)               & 34.7509 & 0.0013                                   & 2.4588 & 0.0008                                            & $-$9.315\tablefootmark{b} & 0.002                  & 29.559 & 0.003\\[0.1cm]
$\omega$                            & (deg)                            & \multicolumn{2}{c}{286$^{\rm +64}_{-\rm 35}$}      & \multicolumn{2}{c}{61$^{\rm +18}_{-\rm 43}$}               & \multicolumn{2}{c}{301$^{\rm +3}_{-\rm 4}$}        & \multicolumn{2}{c}{268$^{\rm +94}_{-\rm 19}$}\\[0.1cm]
$K_{\rm 1}$                         & (m\,s$^{\rm -1}$)                & 24.1    & 1.5                                      & 11.8   & 0.8                                               & \multicolumn{2}{c}{37.3$^{\rm +5.1}_{-\rm 0.9}$}   & 41.3   & 2.9\\[0.1cm]
$Slope$                             & (m\,s$^{\rm -1}$\,yr$^{\rm -1}$) & \multicolumn{2}{c}{...}                            & \multicolumn{2}{c}{...}                                    & 2.629                 & 0.015                      & \multicolumn{2}{c}{...}\\[0.1cm]
$Curvature$                         & (m\,s$^{\rm -1}$\,yr$^{\rm -2}$) & \multicolumn{2}{c}{...}                            & \multicolumn{2}{c}{...}                                    & $-$0.9290             & 0.00012                    & \multicolumn{2}{c}{...}\\[0.1cm]
$f(m)$                              & (10$^{\rm -9}$M$_{\odot}$)       & \multicolumn{2}{c}{3.56$^{\rm +0.71}_{-\rm 0.52}$} & \multicolumn{2}{c}{0.22$^{\rm +0.06}_{-\rm 0.04}$}         & \multicolumn{2}{c}{1.22$^{\rm +0.30}_{-\rm 0.17}$} & \multicolumn{2}{c}{17.6$^{\rm +3.2}_{-\rm 2.3}$}\\[0.1cm]
$a_{\rm 1}\sin i$                   & (10$^{\rm -3}$AU)                & \multicolumn{2}{c}{5.50$^{\rm +0.45}_{-\rm 0.24}$} & \multicolumn{2}{c}{1.41$^{\rm +0.11}_{-\rm 0.09}$}         & \multicolumn{2}{c}{2.25$^{\rm +0.19}_{-\rm 0.13}$} & \multicolumn{2}{c}{9.24$^{\rm +0.61}_{-\rm 0.29}$}\\
\noalign{\vspace{0.05cm}}
\hline
\noalign{\vspace{0.05cm}}
$m_{\rm 2}\sin i$                   & (M$_{\rm Jup}$)                  & \multicolumn{2}{c}{1.56$^{\rm +0.11}_{-\rm 0.10}$} & \multicolumn{2}{c}{0.68$^{\rm +0.07}_{-\rm 0.04}$}         & \multicolumn{2}{c}{1.16$^{\rm +0.10}_{-\rm 0.06}$} & \multicolumn{2}{c}{2.54$^{\rm +0.17}_{-\rm 0.14}$}\\[0.1cm]
$a$                                 & (AU)                             & \multicolumn{2}{c}{3.56$^{\rm +0.24}_{-\rm 0.15}$} & \multicolumn{2}{c}{2.45$^{\rm +0.04}_{-\rm 0.06}$}         & \multicolumn{2}{c}{2.15$^{\rm +0.04}_{-\rm 0.05}$} & \multicolumn{2}{c}{3.43$^{\rm +0.20}_{-\rm 0.12}$}\\
\noalign{\vspace{0.05cm}}
\hline
\noalign{\vspace{0.05cm}}
$N_{\rm mes}$                       &                                  & \multicolumn{2}{c}{22}                             & \multicolumn{2}{c}{45}                                     & \multicolumn{2}{c}{32}                             & \multicolumn{2}{c}{18}\\[0.1cm]
${\rm Med} (t_{\rm exp})$           & (s)                              & \multicolumn{2}{c}{158}                            & \multicolumn{2}{c}{106}                                    & \multicolumn{2}{c}{60}                             & \multicolumn{2}{c}{173}\\[0.1mm]
$\langle {\rm S/N} \rangle$         &                                  & \multicolumn{2}{c}{52}                             & \multicolumn{2}{c}{58}                                     & \multicolumn{2}{c}{51}                             & \multicolumn{2}{c}{45}\\[0.1mm]
$\langle \epsilon_{\rm RV} \rangle$ & (m\,s$^{\rm -1}$)                & \multicolumn{2}{c}{1.59}                           & \multicolumn{2}{c}{1.57}                                   & \multicolumn{2}{c}{1.34}                           & \multicolumn{2}{c}{1.73}\\[0.1cm]
$Span$                              & (d)                              & \multicolumn{2}{c}{2028}                           & \multicolumn{2}{c}{2220}                                   & \multicolumn{2}{c}{1989}                           & \multicolumn{2}{c}{1986}\\[0.1cm]
$\sigma_{\rm O-C}$                  & (m\,s$^{\rm -1}$)                & \multicolumn{2}{c}{1.79}                           & \multicolumn{2}{c}{2.49}                                   & \multicolumn{2}{c}{2.19}                           & \multicolumn{2}{c}{1.60}\\[0.1cm]
$\chi^{\rm^ 2}_{\rm red}$           &                                  & \multicolumn{2}{c}{1.50}                           & \multicolumn{2}{c}{2.48}                                   & \multicolumn{2}{c}{3.08}                           & \multicolumn{2}{c}{1.22}\\[0.1cm]
\hline
\noalign{\vspace{0.05cm}}
{\bf Parameter}                     & {\bf Unit}                       & \multicolumn{2}{c}{\bf HD\,43197}                  & \multicolumn{2}{c}{\bf HD\,44219}                         & \multicolumn{2}{c}{\bf HD\,148156}                         & \multicolumn{2}{c}{\bf HD\,156411}\\
\hline
\noalign{\vspace{0.05cm}}
$P$                                 & (d)                              & 327.8 & 1.2                                        & \multicolumn{2}{c}{472.3$^{\rm+6.3}_{-\rm 5.0}$}          & 1027   & 28                                                & 842.2      & 14.5\\[0.1cm]
$T$                                 & (JD\tablefootmark{a})            & 713.8 & 1.7                                        & \multicolumn{2}{c}{585.6$^{\rm+7.2}_{-\rm 3.1}$}          & 707    & 20                                                & \multicolumn{2}{c}{356$^{\rm +46}_{-\rm 33}$}\\[0.1cm]
$e$                                 &                                  & \multicolumn{2}{c}{0.83$^{\rm+0.05}_{-\rm 0.01}$}  & \multicolumn{2}{c}{0.61$^{\rm+0.07}_{-\rm 0.09}$}         & \multicolumn{2}{c}{0.52$^{\rm+0.04}_{-\rm 0.09}$}          & 0.22       & 0.08\\[0.1cm]
$\gamma$                            & (km\,s$^{\rm -1}$)               & \multicolumn{2}{c}{72.512$^{\rm +4}_{-\rm 11}$}    & $-$12.0732\tablefootmark{c} & 0.0026                      & \multicolumn{2}{c}{$-$1.7480$^{\rm+0.0013}_{-\rm 0.0009}$} & $-$38.7512 & 0.0011\\[0.1cm]
$\omega$                            & (deg)                            & \multicolumn{2}{c}{251$^{\rm +7}_{-\rm 12}$}       & \multicolumn{2}{c}{147.4$^{\rm +6.2}_{-\rm 3.2}$}         & \multicolumn{2}{c}{35$^{\rm +17}_{-\rm 12}$}               & \multicolumn{2}{c}{45$^{\rm +22}_{-\rm 14}$}\\[0.1cm]
$K_{\rm 1}$                         & (m\,s$^{\rm -1}$)                & \multicolumn{2}{c}{32.4$^{\rm +10.9}_{-\rm 3.3}$}  & 19.4                  & 3.0                               & \multicolumn{2}{c}{17.5$^{\rm +1.5}_{-\rm 0.5}$}           & 14.0       & 0.8\\[0.1cm]
$Slope$                             & (m\,s$^{\rm -1}$\,yr$^{\rm -1}$) & \multicolumn{2}{c}{...}                            & $-$0.637              & 0.016                             & \multicolumn{2}{c}{...}                                    & \multicolumn{2}{c}{...}\\[0.1cm]
$Curvature$                         & (m\,s$^{\rm -1}$\,yr$^{\rm -2}$) & \multicolumn{2}{c}{...}                            & $-$0.76390            & 0.000                             & \multicolumn{2}{c}{...}                                    & \multicolumn{2}{c}{...}\\[0.1cm]
$f(m)$                              & (10$^{\rm -9}$M$_{\odot}$)       & \multicolumn{2}{c}{0.21$^{\rm +0.13}_{-\rm 0.04}$} & \multicolumn{2}{c}{0.17$^{\rm +0.05}_{-\rm 0.03}$}        & \multicolumn{2}{c}{0.35$^{\rm +0.08}_{-\rm 0.04}$}         & \multicolumn{2}{c}{0.23$^{\rm +0.04}_{-\rm 0.03}$}\\[0.1cm]
$a_{\rm 1}\sin i$                   & (10$^{\rm -3}$AU)                & \multicolumn{2}{c}{0.55$^{\rm +0.10}_{-\rm 0.04}$} & \multicolumn{2}{c}{0.66$^{\rm +0.06}_{-\rm 0.04}$}        & \multicolumn{2}{c}{1.41$^{\rm +0.09}_{-\rm 0.06}$}         & \multicolumn{2}{c}{1.06$^{\rm +0.08}_{-\rm 0.05}$}\\
\noalign{\vspace{0.05cm}}
\hline
\noalign{\vspace{0.05cm}}
$m_{\rm 2}\sin i$                   & (M$_{\rm Jup}$)                  & \multicolumn{2}{c}{0.60$^{\rm +0.12}_{-\rm 0.04}$} & \multicolumn{2}{c}{0.58$^{\rm +0.06}_{-\rm 0.04}$}        & \multicolumn{2}{c}{0.85$^{\rm +0.67}_{-\rm 0.05}$}         & \multicolumn{2}{c}{0.74$^{\rm +0.05}_{-\rm 0.04}$}\\[0.1cm]
$a$                                 & (AU)                             & \multicolumn{2}{c}{0.92$^{\rm +0.01}_{-\rm 0.02}$} & 1.19                  & 0.02                              & \multicolumn{2}{c}{2.45$^{\rm +0.04}_{-\rm 0.05}$}         & \multicolumn{2}{c}{1.88$^{\rm +0.03}_{-\rm 0.04}$}\\
\noalign{\vspace{0.05cm}}
\hline
\noalign{\vspace{0.05cm}}
$N_{\rm mes}$                       &                                  & \multicolumn{2}{c}{33}                             & \multicolumn{2}{c}{46}                                    & \multicolumn{2}{c}{42}                                     & \multicolumn{2}{c}{50}\\[0.1cm]
${\rm Med} (t_{\rm exp})$           & (s)                              & \multicolumn{2}{c}{171}                            & \multicolumn{2}{c}{63}                                    & \multicolumn{2}{c}{92}                                     & \multicolumn{2}{c}{90}\\[0.1mm]
$\langle {\rm S/N} \rangle$         &                                  & \multicolumn{2}{c}{48}                             & \multicolumn{2}{c}{51}                                    & \multicolumn{2}{c}{56}                                     & \multicolumn{2}{c}{73}\\[0.1mm]
$\langle \epsilon_{\rm RV} \rangle$ & (m\,s$^{\rm -1}$)                & \multicolumn{2}{c}{1.42}                           & \multicolumn{2}{c}{1.55}                                  & \multicolumn{2}{c}{2.24}                                   & \multicolumn{2}{c}{1.26}\\[0.1cm]
$Span$                              & (d)                              & \multicolumn{2}{c}{1943}                           & \multicolumn{2}{c}{1988}                                  & \multicolumn{2}{c}{2168}                                   & \multicolumn{2}{c}{2231}\\[0.1cm]
$\sigma_{\rm O-C}$                  & (m\,s$^{\rm -1}$)                & \multicolumn{2}{c}{1.44}                           & \multicolumn{2}{c}{2.39}                                  & \multicolumn{2}{c}{3.69}                                   & \multicolumn{2}{c}{2.94}\\[0.1cm]
$\chi^{\rm^ 2}_{\rm red}$           &                                  & \multicolumn{2}{c}{1.19}                           & \multicolumn{2}{c}{2.62}                                  & \multicolumn{2}{c}{2.89}                                   & \multicolumn{2}{c}{5.21}\\[0.1cm]
\hline
\end{tabular}
\tablefoot{Errobars are Monte-Carlo based 1$\sigma$ uncertainties. 
$\langle \epsilon_{\rm RV} \rangle$ is the radial-velocity uncertainty weighted average. 
$\sigma_{\rm O-C}$ is the weighted {\it rms} of the residuals to the fitted orbits. 
${\rm Med} (t_{\rm exp})$ is the median exposure time and $\langle {\rm S/N} \rangle$ is the 
average signal-to-noise ratio at 550\,nm.\\
\tablefoottext{a}{JD\,=\,BJD$-$2\,454\,000}
\tablefoottext{b}{at BJD\,=2\,454\,106.6}
\tablefoottext{c}{at BJD\,=2\,454\,512.6}}
\end{table*}

\begin{figure*}
\centering
\begin{tabular}{ccc}
\includegraphics[width=0.3\hsize]{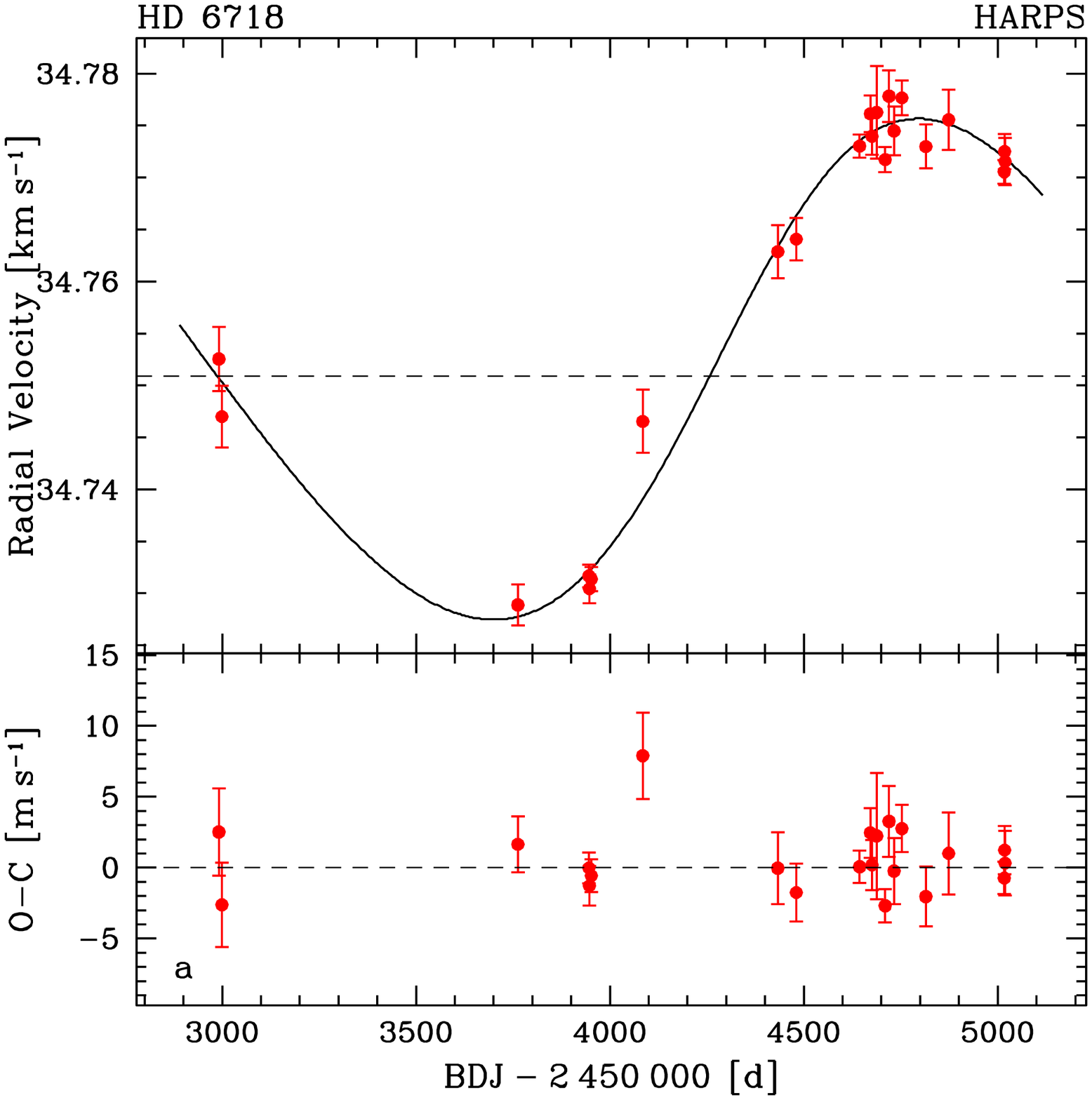} & \includegraphics[width=0.3\hsize]{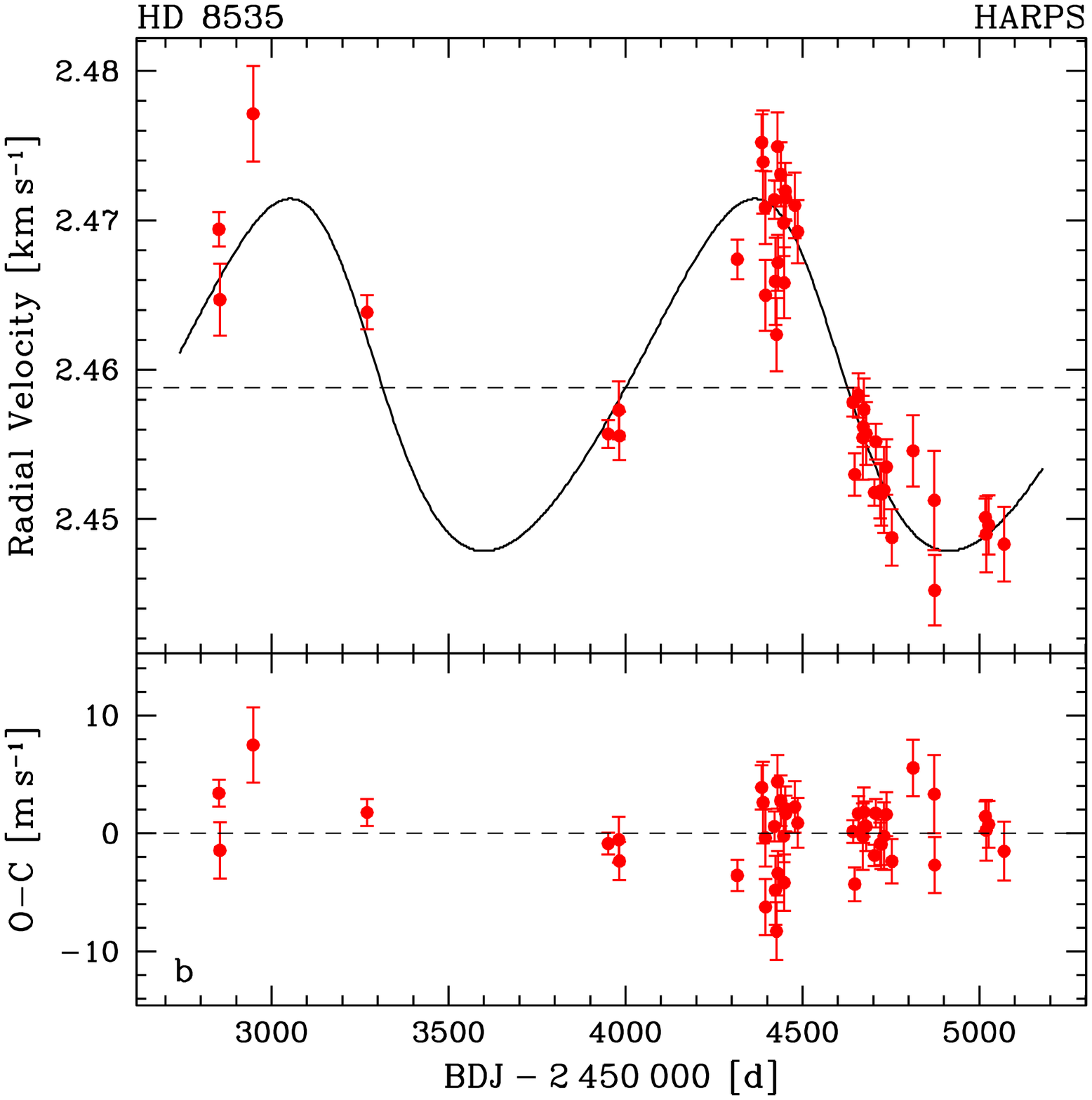} & \includegraphics[width=0.3\hsize]{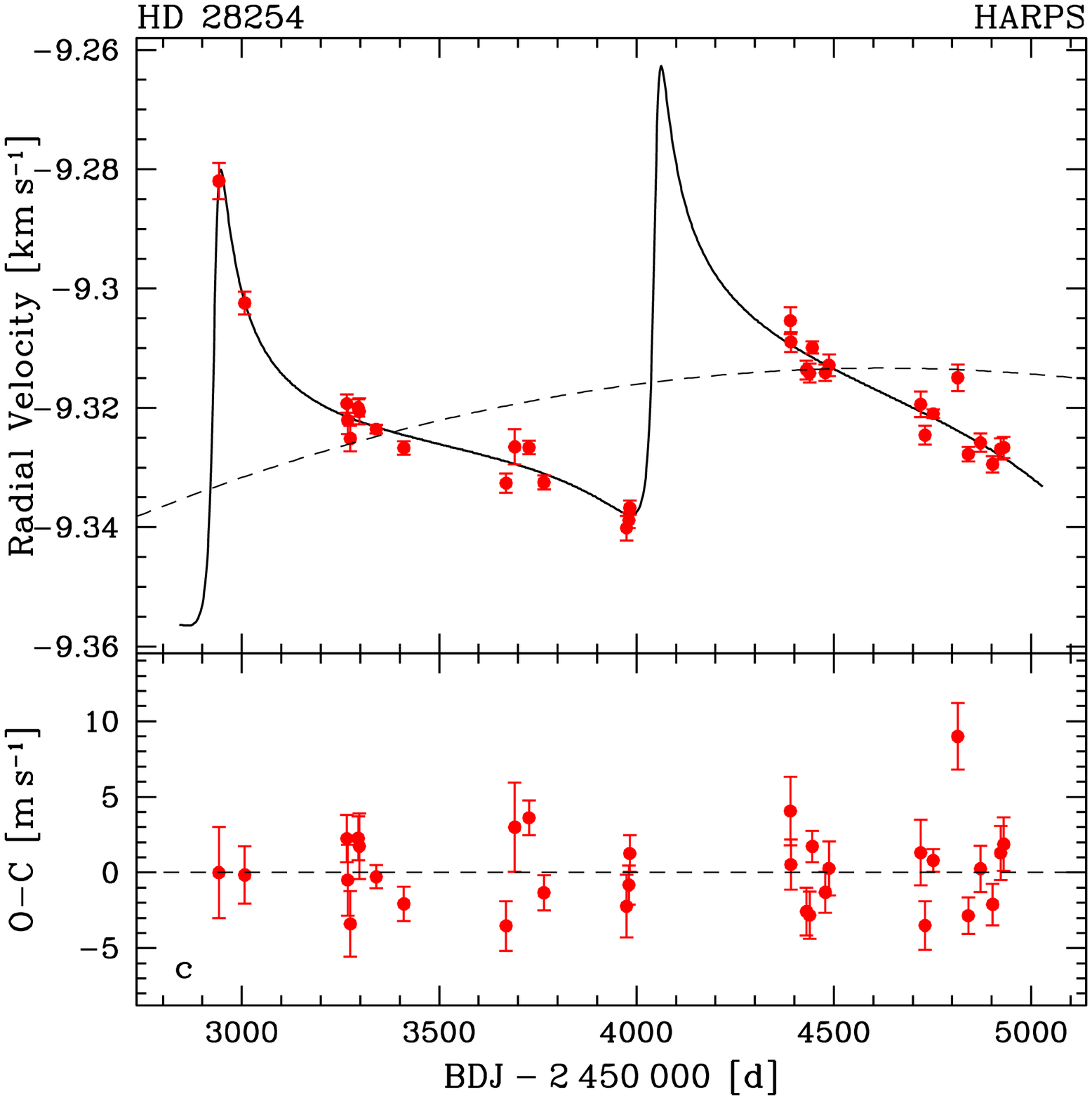}\\
\includegraphics[width=0.3\hsize]{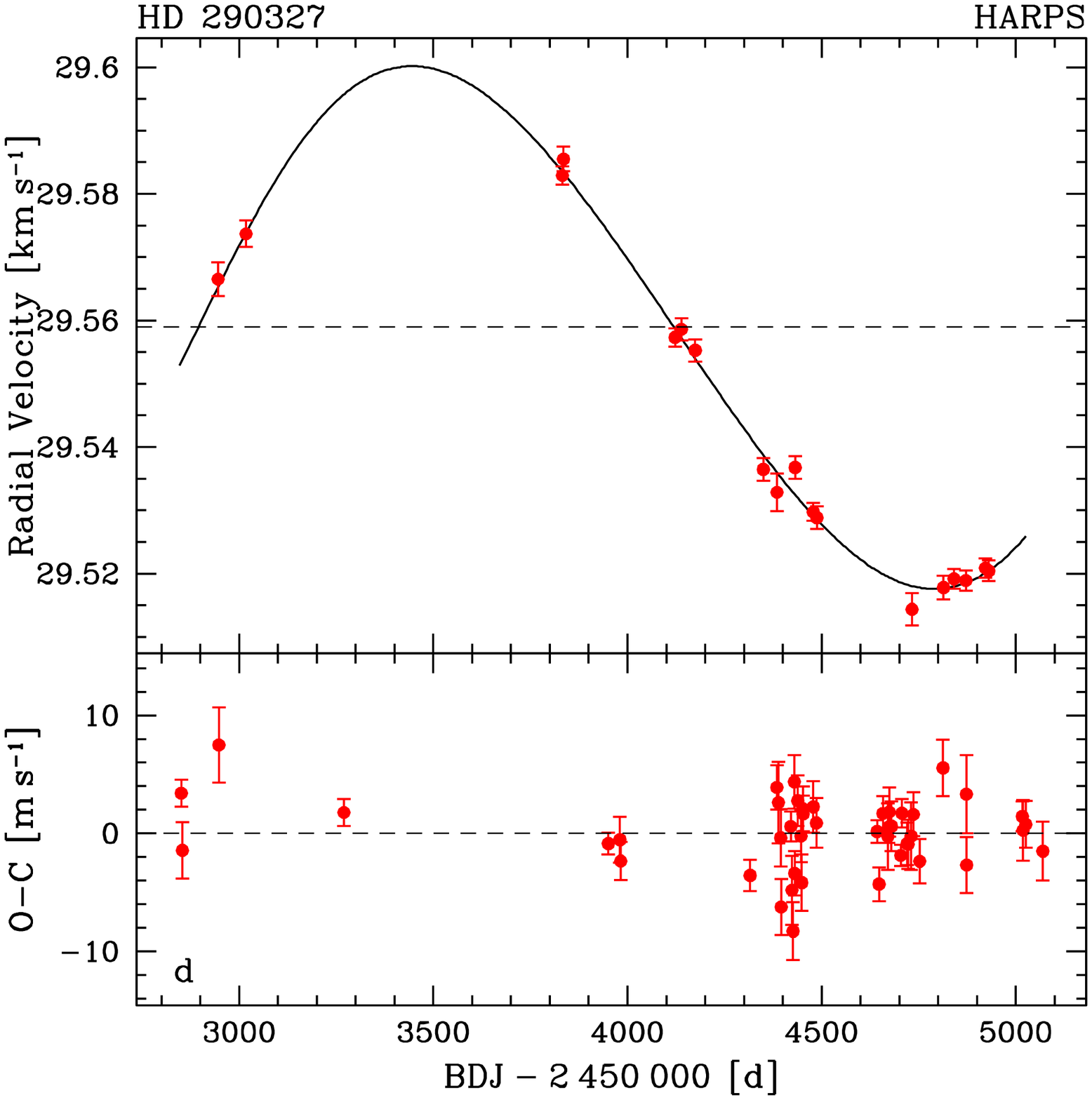} & \includegraphics[width=0.3\hsize]{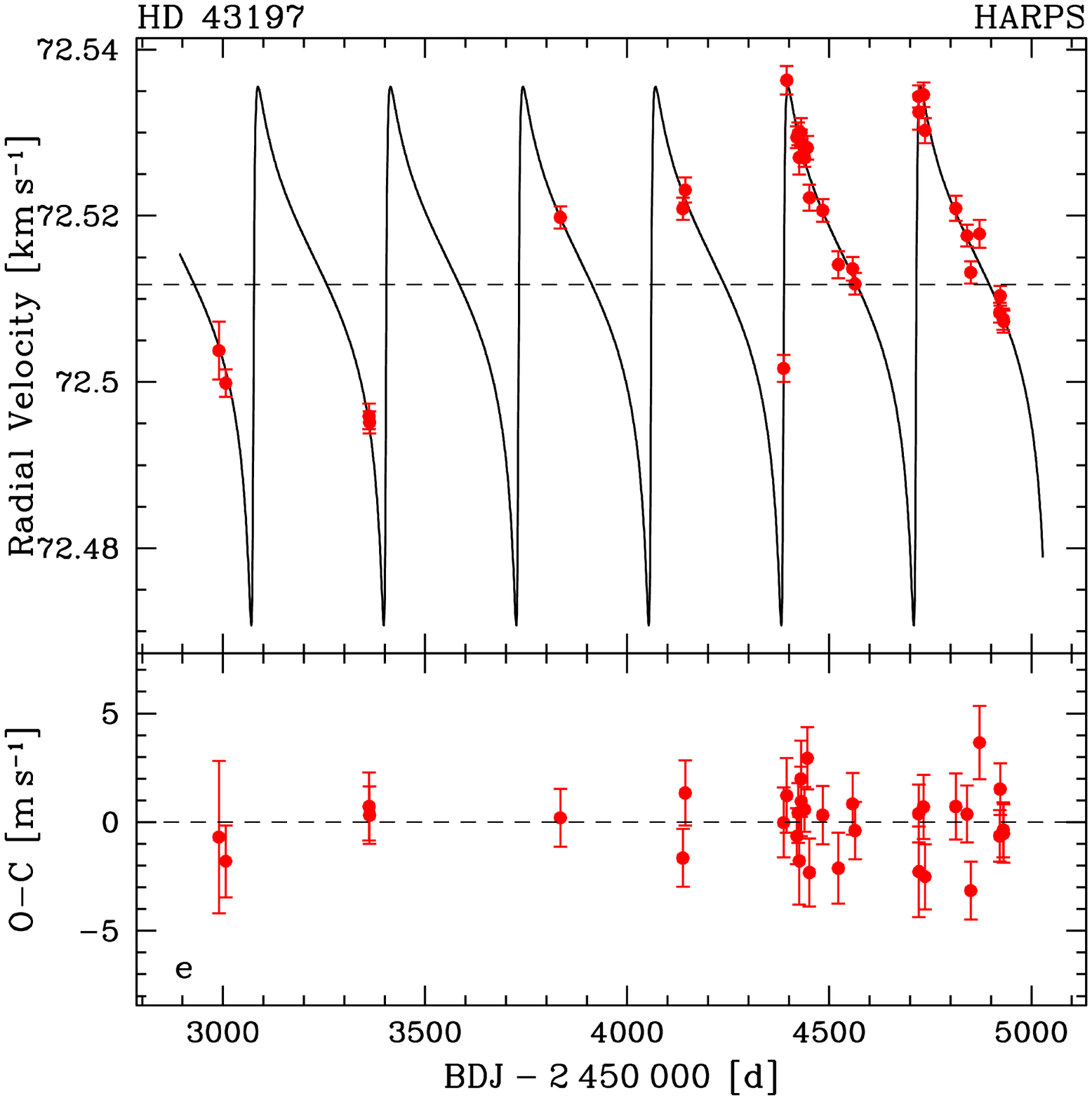} & \includegraphics[width=0.3\hsize]{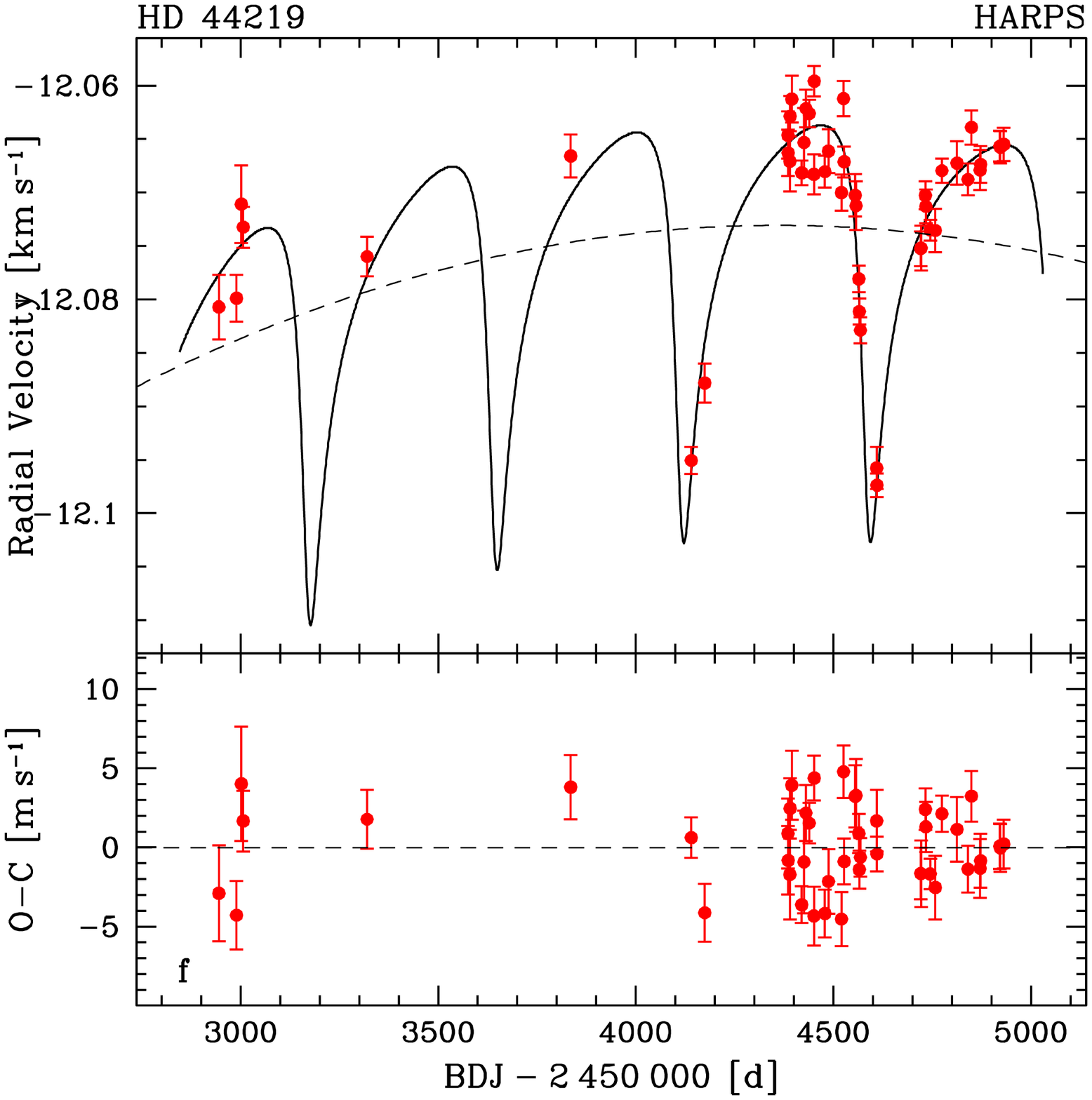}\\
\includegraphics[width=0.3\hsize]{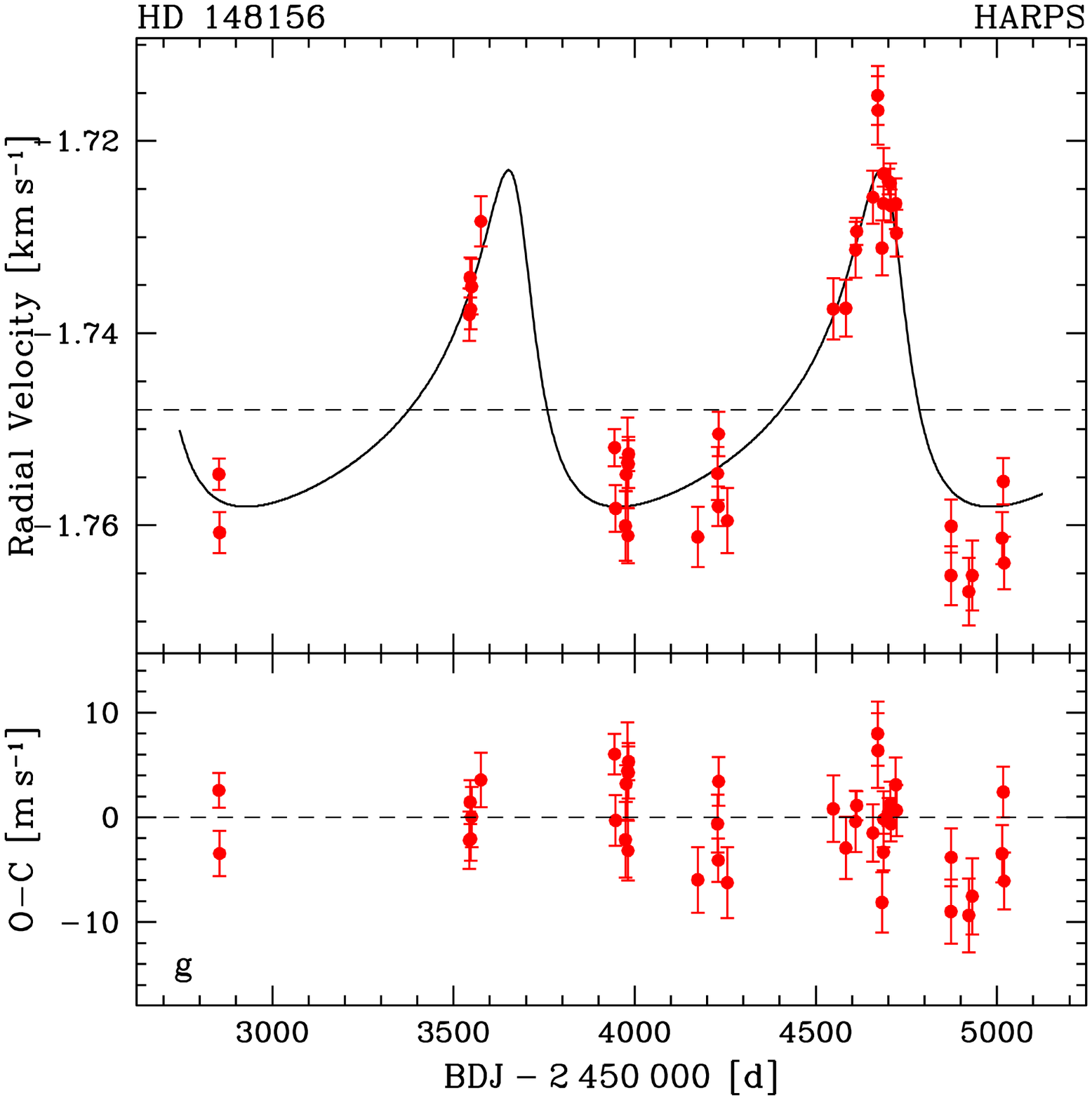} &                                                  & \includegraphics[width=0.3\hsize]{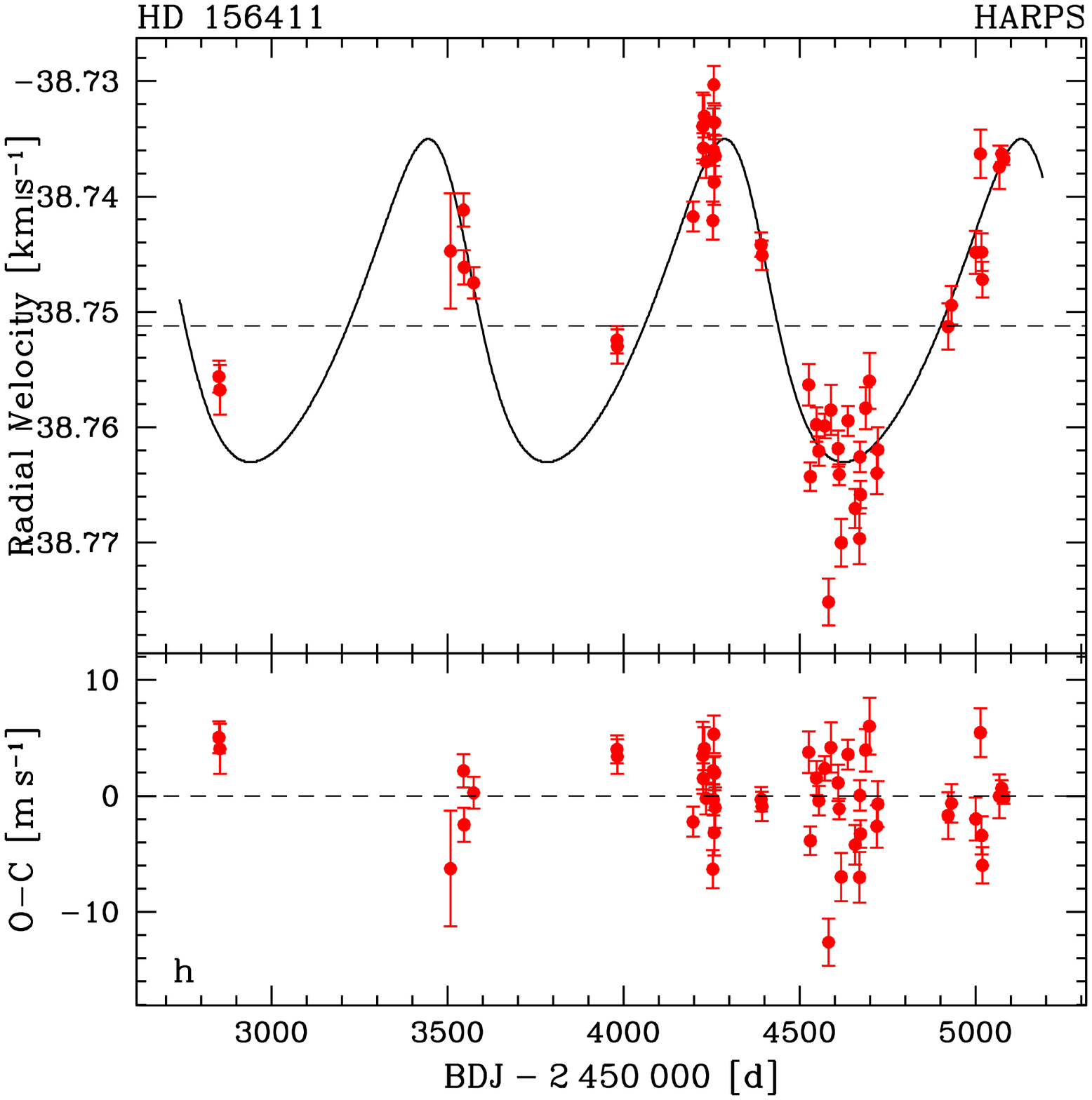} \\
\end{tabular}
\caption{{\footnotesize HARPS} radial-velocity data for the eight stars presented in this paper. Temporal velocities (filled dots) and 
fitted orbital solutions (solid lines) are displayed in the upper panels. Residuals to the fitted orbits are displayed in the bottom 
panels. For \object{{\footnotesize HD}\,28254} and \object{{\footnotesize HD}\,44219}, the dashed lines in the upper panels show 
additional quadratic radial-velocity drifts that were included to the orbital fits. For all the other targets, the upper panels dashed 
lines represent the systemic velocity $\gamma$.}
\label{orbits}
\end{figure*}

\subsection{HD\,6718}\label{hd6718}

The 22 {\footnotesize HARPS} radial-velocity measurements were gathered for \object{{\footnotesize HD}\,6718} ({\footnotesize HIP}\,5301) 
between {\footnotesize BJD}\,=\,2\,452\,991 (December 18, 2003) and {\footnotesize BJD}\,=\,2\,455\,019 (July 6, 2009). These velocities 
have a mean uncertainty of 1.59\,m\,s$^{\rm -1}$. We have fitted a long-period Keplerian orbit to these data. The orbital parameters 
resulting from this fit are listed in Table~\ref{rvsol}. With these parameters and assuming a mass of 0.96\,M$_\odot$ for the host star, 
we computed a minimum mass of 1.56\,M$_{\rm Jup}$ for the companion and a semi-major axis of 3.56\,AU. Residuals to the fitted orbit have 
a very low dispersion ($\sigma_{\rm O-C}$\,=\,1.79\,m\,s$^{\rm -1}$), and a 1-planet model is thus sufficient for explaining the detected 
signal. We display our data and the fitted orbit in panel {\bf a} of Fig.~\ref{orbits}.

\begin{figure*}[t!]
\centering
\begin{tabular}{ccc}
\includegraphics[width=0.3\hsize]{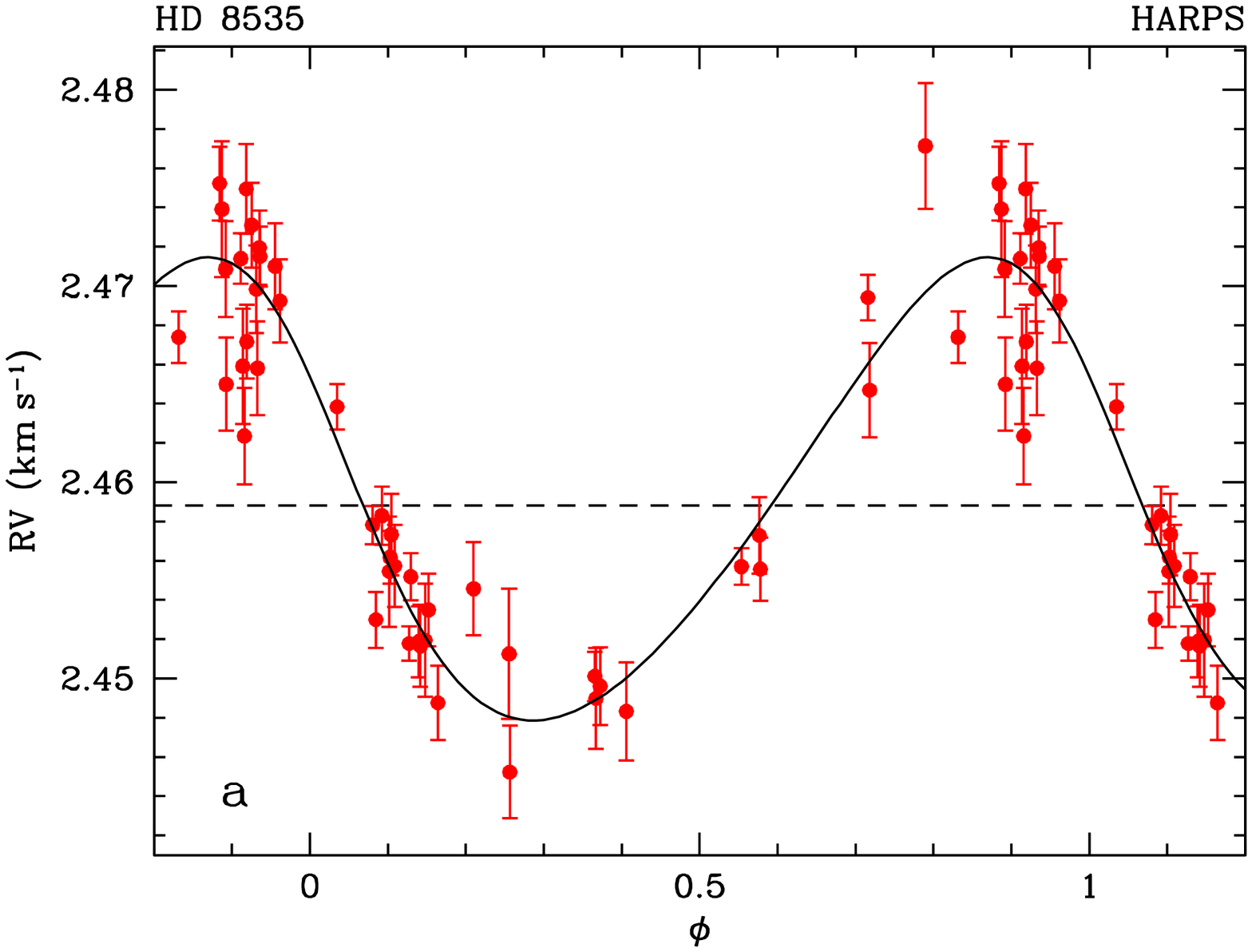} & \includegraphics[width=0.3\hsize]{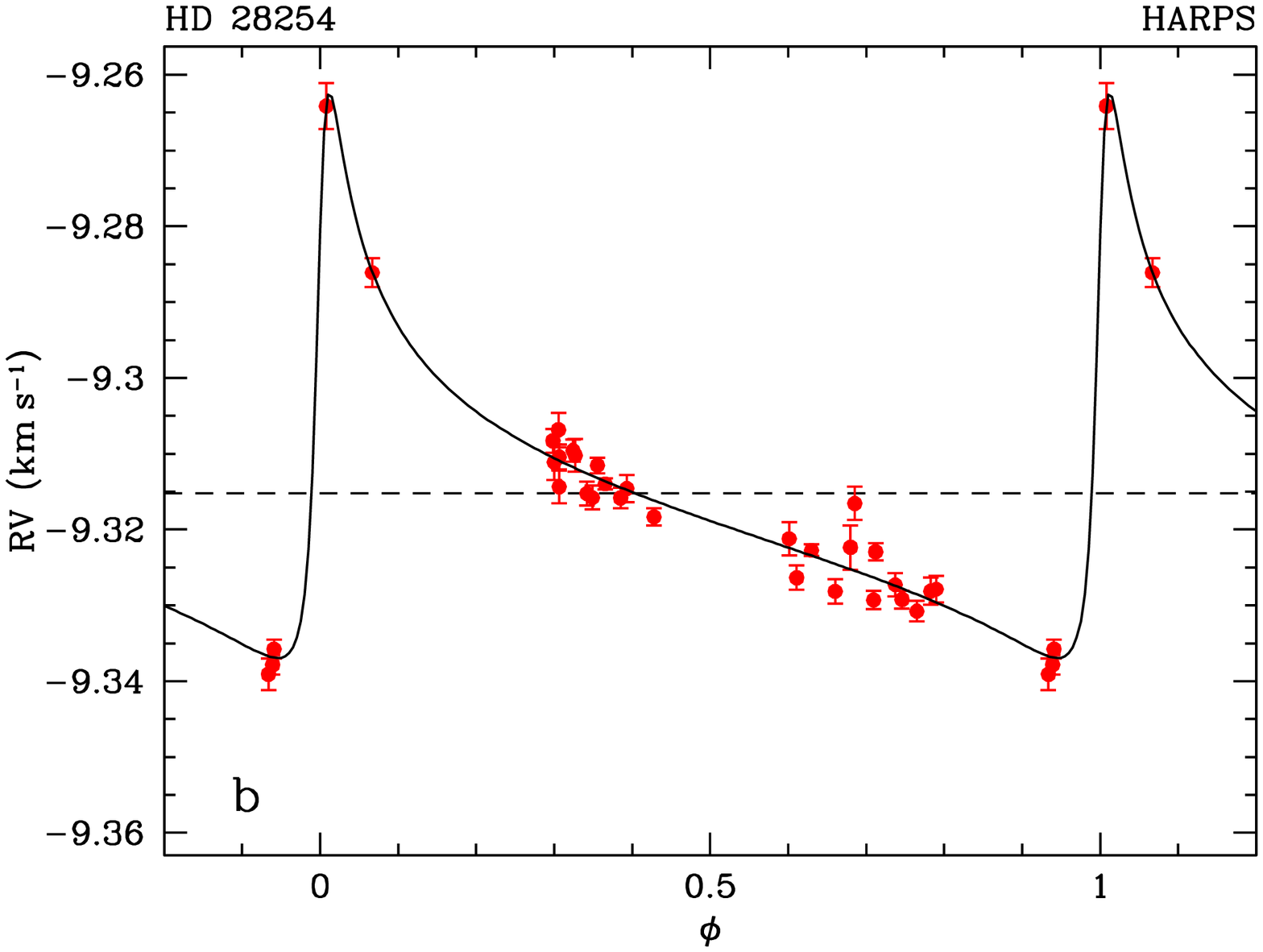} & \includegraphics[width=0.3\hsize]{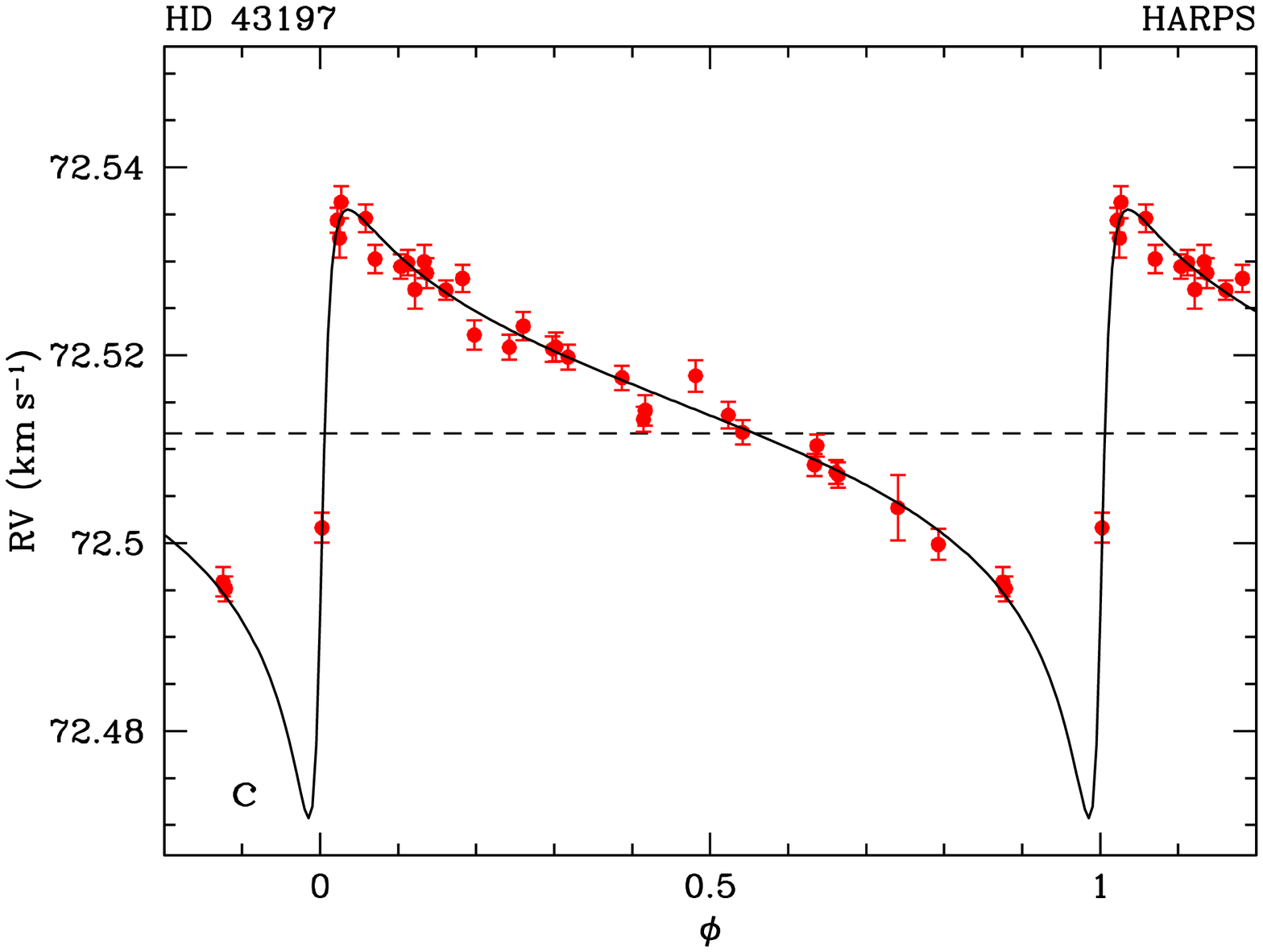}\\
\includegraphics[width=0.3\hsize]{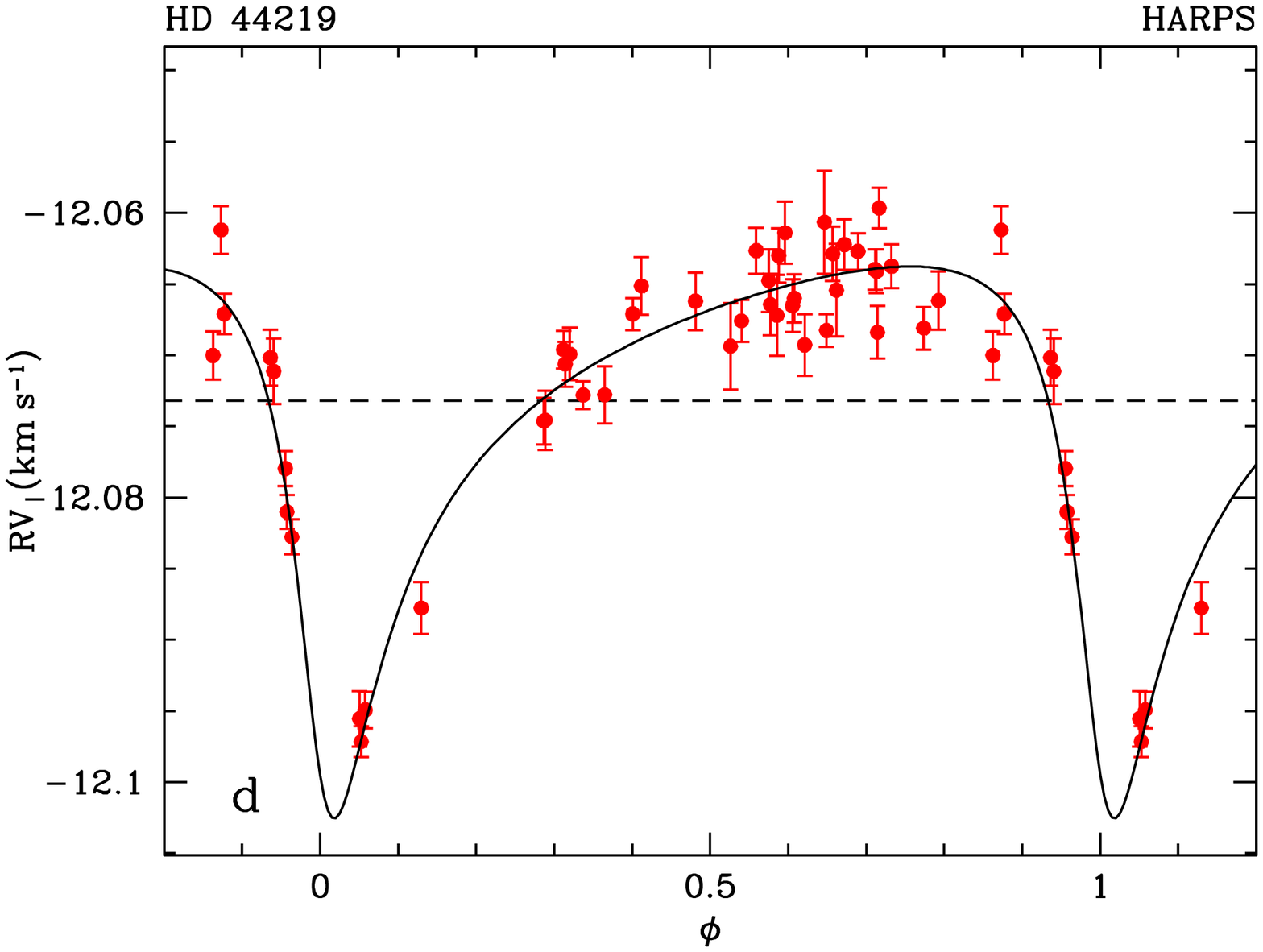} & \includegraphics[width=0.3\hsize]{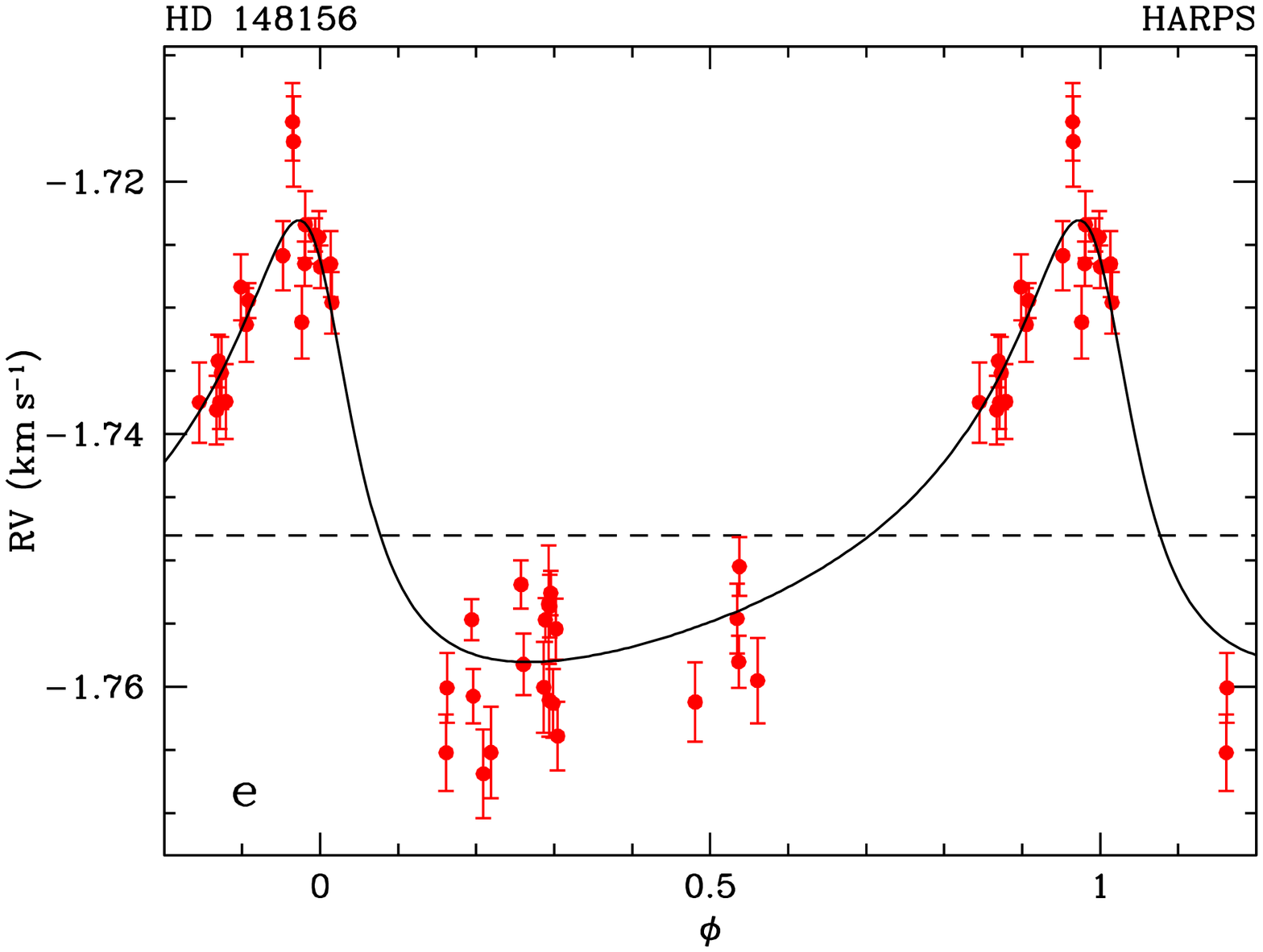} & \includegraphics[width=0.3\hsize]{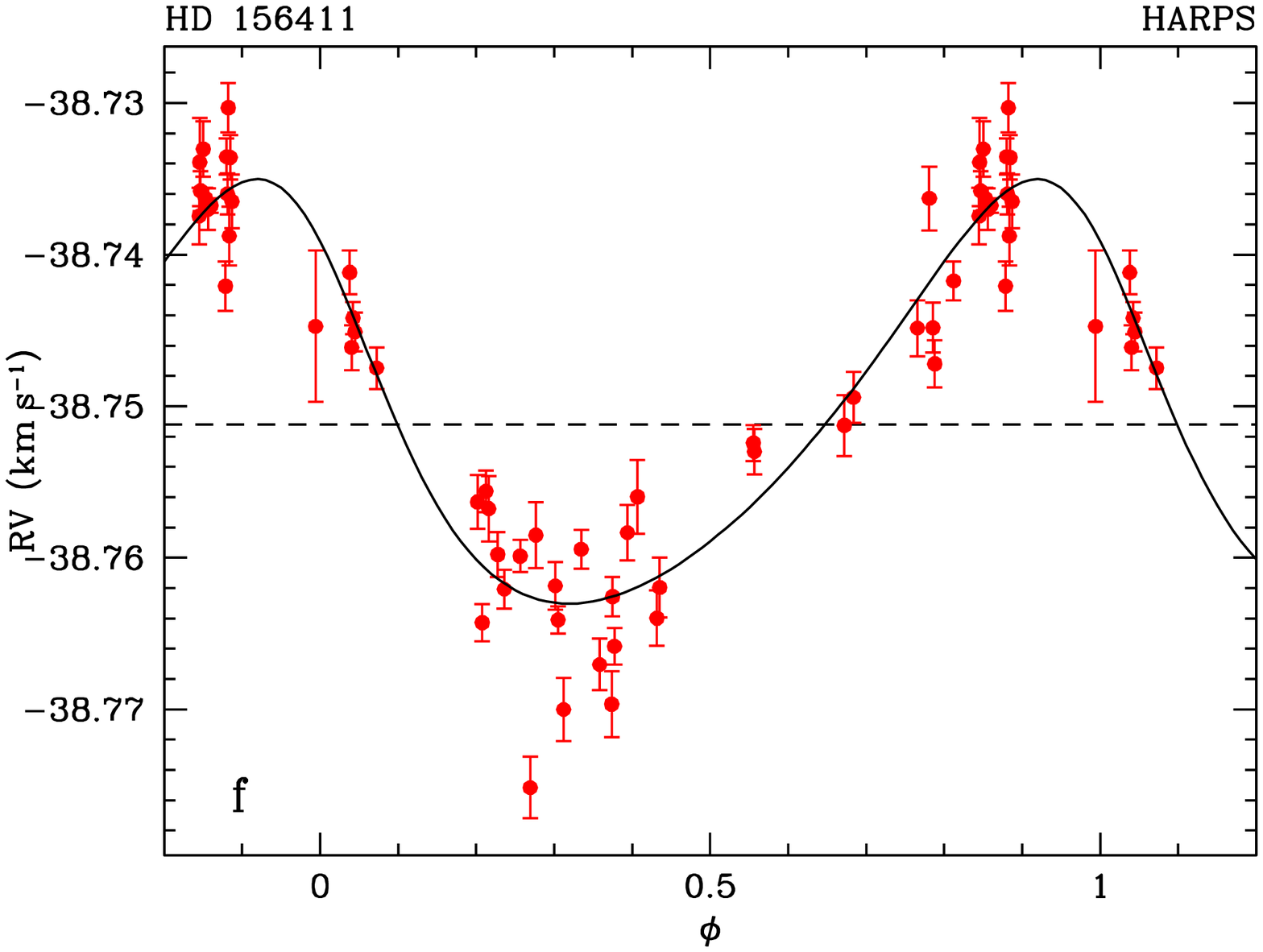}\\
\end{tabular}
\caption{{\footnotesize HARPS} phase-folded velocities and fitted orbits for the 6 targets observed for at least two orbital periods. The 
radial-velocities of \object{{\footnotesize HD}\,28254} and \object{{\footnotesize HD}\,44219} have been corrected for the fitted 
quadratic drift.}
\label{phased}
\end{figure*}

\subsection{HD\,8535}\label{hd8535}

A set of 45 {\footnotesize HARPS} radial velocities was obtained for \object{{\footnotesize HD}\,8535} ({\footnotesize HIP}\,6511) between 
{\footnotesize BJD}\,=\,2\,452\,850 (July 29, 2003) and {\footnotesize BJD}\,=\,2\,455\,070 (August 26, 2009). The mean radial-velocity 
uncertainty of these measurements is 1.57\,m\,s$^{\rm -1}$. The presence of a long-period signal is obvious (see panel~{\bf b} of 
Fig.~\ref{orbits}). The results of the Keplerian fit to our data are presented in Table~\ref{rvsol}. The dispersion of the residuals to 
the fitted orbit are somewhat larger than expected ($\chi^{\rm2}_{red}$\,=\,2.48). We have tried to fit orbits including additional 
signals (linear or quadratic trend; 2-Keplerian model) but we were unable to significantly improve the quality of the solution. We have 
also verified the absence of any correlation between the residuals and the {\footnotesize CCF} bisectors.

Assuming a primary mass of 1.13\,M$_{\odot}$, we computed the minimum mass of the companion and the semi-major axis of its orbit: 
$m_{\rm 2}sin i$\,=\,0.68\,M$_{\rm Jup}$ and $a$\,=\,2.45\,AU. The fitted orbit and the fit residuals are displayed in panel~{\bf b} of 
Fig.~\ref{orbits}. The phase-folded orbit and velocities are displayed in panel {\bf a} of Fig.~\ref{phased}.

\subsection{HD\,28254}\label{hd28254}

We collected 32 {\footnotesize HARPS} radial velocities for \object{{\footnotesize HD}\,28254} ({\footnotesize HIP}\,20606) between 
{\footnotesize BJD}\,=\,2\,452\,941 (October 28, 2003) and {\footnotesize BJD}\,=\,2\,454\,931 (April 9, 2009). The average velocity 
uncertainty of these measurements is 1.34\,m\,s$^{\rm -1}$. As indicated in Sect.~\ref{stars}, \object{{\footnotesize HD}\,28254} is the 
main component of  a visual binary system. The presence of a nearby visual companion can induce radial-velocity errors via the 
seeing-dependent contamination of the spectrum of the main component. \citet{PepeIAU} have simulated this effect for {\footnotesize HARPS} 
and using the La Silla seeing statisitics. We have performed the same simulations using the characteristics of the 
\object{{\footnotesize HD}\,28254} system ($\rho$\,=\,4.3\arcsec, $\Delta m_{\rm V}$\,=\,6.11). The expected effect is negligible in this 
case ($\ll$0.1\,m\,s$^{\rm -1}$).

A long-period signal is visible on the data (see panel {\bf c} of Fig.~\ref{orbits}). From the visual inspection of these data, it is also 
obvious that the Keplerian signal is highly eccentric. We first attempted to fit a single orbit to our velocities. With $e$\,$>$\,0.99, 
the obtained orbit was not physical so we rejected it. After this initial unsuccessful attempt, we tried to fit orbits with additional 
signals. The following two models did not give satisfactory results: the 1-Keplerian plus linear drift model and the 2-Keplerian one. The 
best fit we obtained and finally adopted is based on a 1-Keplerian plus quadratic drift model. Despite the poor coverage of the second 
radial-velocity maximum, hence the poorly constrained velocity semi-amplitude, the planetary nature of {\footnotesize HD}\,23254\,b is 
well established. The orbital period is very well constrained. Solutions with significantly larger amplitudes are possible, but they also 
have much larger eccentricities. The effect of the increasing amplitude on the minimum mass of the companion is largely compensated for 
by the increasing eccentricity as the mass function scales with $K_{\rm 1}^{\rm 3}(1-e^{\rm 2})^{\rm 3/2}$. 

More data are necessary for better constraining the shape of the orbit of \object{{\footnotesize HD}\,28254}\,b. In particular, a better 
coverage of the periastron passage will allow us to constrain the radial-velocity semi-amplitude and orbital eccentricity. These future 
observations will also help us characterizing the secondary signal better. The orbital solution we present here is thus a preliminary one. 
The observed additional quadratic signal remains compatible with the signal that the visual companion described in Sect.~\ref{stars} could 
induce.

Using our fitted parameters, we computed the minimum mass of the planetary companion and its orbital semi-major axis: 
$m_{\rm 2}sin i$\,=\,1.16\,M$_{\rm Jup}$ and  $a$\,=\,2.15\,AU. Our radial-velocity data and the fitted orbit are displayed in panel 
{\bf c} of Fig.~\ref{orbits}. The phase-folded orbit and velocities are displayed in panel {\bf b} of Fig.~\ref{phased}.

\subsection{HD\,290327}\label{hd290327}

Between {\footnotesize BJD}\,=\,2\,452\,945 (November 1, 2003) and {\footnotesize BJD}\,=\,2\,454\,931 (April 9, 2009), we have observed 
{\footnotesize HD}\,290327 ({\footnotesize HIP}\,25191) 18 times with {\footnotesize HARPS}. The obtained radial velocities have a mean 
uncertainty of 1.73\,m\,s$^{\rm -1}$. A long-period signal is well visible on the data. We list in Table~\ref{rvsol} the parameters 
resulting from a Keplerian fit to the data. With these parameters and using the primary mass of Table~\ref{tabstars} (0.90\,M$_\odot$), 
we computed the minimum mass of the companion that induces the detected radial-velocity signal and the semi-major axis of its orbit: 
$m_{\rm 2}sin i$\,=\,2.54\,M$_{\rm Jup}$ and $a$\,=\,3.43\,AU. The dispersion of the residuals to the fitted orbit is very low: 
1.6\,m\,s$^{\rm -1}$. Our velocity data and the fitted orbit are displayed in panel {\bf d} of Fig.~\ref{orbits}.

\subsection{HD\,43197}\label{hd43197}

We have gathered a set of 33 radial velocities for \object{{\footnotesize HD}\,43197} ({\footnotesize HIP}\,29550). These data were 
obtained between {\footnotesize BJD}\,=\,2\,452\,989 (December 15, 2003) and {\footnotesize BJD}\,=\,2\,454\,932 (April 10, 2009), and 
they have a mean velocity uncertainty of 1.42\,m\,s$^{\rm -1}$. A visual inspection of the data displayed in panel~{\bf e} of 
Fig.~\ref{orbits} shows  a periodic highly eccentric signal. The parameters resulting from a Keplerian fit to the data are listed in 
Table~\ref{rvsol}. The fitted eccentricity is indeed very high and very well constrained in spite of our poor coverage of the 
radial-velocity minimum. The {\it rms} of the residuals to the fitted orbit is very low: 1.42\,m\,s$^{\rm -1}$. This 1-planet fit is thus 
satisfactory. With a primary mass of 0.96\,M$_\odot$, the computed minimum mass of the companion is 0.6\,M$_{\rm Jup}$. The orbital 
semi-major axis is 0.92\,AU. Our data, the fitted orbit, and the residuals are displayed in panel {\bf e} of Fig.~\ref{orbits}. Panel 
{\bf c} of Fig.~\ref{phased} shows the phase-folded orbit.

\subsection{HD\,44219}\label{hd44219}

The {\footnotesize HARPS} data set we have in hand for \object{{\footnotesize HD}\,44219} ({\footnotesize HIP}\,30114) consists of 46 
radial velocities measured between {\footnotesize BJD}\,=\,2\,452\,944 (October 31, 2003) and {\footnotesize BJD}\,=\,2\,454\,932 (April 
10, 2009). These velocities have an average uncertainty of 1.55\,m\,s$^{\rm -1}$. We first fitted a 1-Keplerian model to these data. The 
residuals to this fitted orbit were a bit large ($\sigma_{\rm O-C}$\,=\,2.95\,m\,s$^{\rm -1}$). We then tried to include an additional 
signal in the fitted model. Adding a linear drift improved the quality of the fit in a marginally significant way 
($\chi^{\rm 2}_{\rm red}$ decreasing from 3.79 to 3.30; f-test probability\,=\,71\%). The improvement resulting from including a quadratic 
trend in the fitted model was more significant: $\chi^{\rm 2}_{\rm red}$\,=\,2.62 and f-test probability equals 90\%. We also tried to fit 
a 2-planet model but we were unable to find any convincing solution. We finally decided to adopt the Keplerian\,+\,quadratic drift model. 
The parameters resulting from this fit are listed in Table~\ref{rvsol}. The orbital parameters obtained with our initial 1-planet fit did 
not differ much from the ones in the adopted solution ($P$\,=447\,d; $e$\,=\,0.57; $K_{\rm 1}$\,=\,17\,m\,s$^{\rm -1}$ and 
$m_{\rm 2}\sin i$\,=\,0.52\,M$_{\rm Jup}$).

The dispersion of the residuals to the finally selected solution is still a bit large ($\sigma_{\rm O-C}$\,=\,2.39\,m\,s$^{\rm -1}$), but 
this is probably due to the still not very well constrained additional signal. With this solution and assuming $M_{*}$\,=\,1\,M$_{\odot}$, 
the computed minimum mass for \object{{\footnotesize HD}\,44219}\,b is 0.58\,M$_{\rm Jup}$ and its orbital semi-major axis is 1.19\,AU. We 
display the fitted model and the temporal residuals in panel~{\bf f} of Fig.~\ref{orbits}. The phase-folded orbit and velocities are 
displayed in panel {\bf d} of Fig.~\ref{phased}.

\subsection{HD\,148156}\label{hd148156}

We have collected 42 radial velocities with {\footnotesize HARPS} for \object{{\footnotesize HD}\,148156} ({\footnotesize HIP}\,80680) 
between {\footnotesize BJD}\,=\,2\,452\,852 (July 31, 2003) and {\footnotesize BJD}\,=\,2\,455\,020 (July 7, 2009). The mean 
radial-velocity uncertainty of this data set is 2.24\,m\,s$^{\rm -1}$. A long-period signal is visible in the data displayed in 
panel~{\bf g} of Fig.~\ref{orbits}. The results of the Keplerian fit performed on our velocity set are presented in Table~\ref{rvsol}. 
The {\it rms} of the residuals is a bit large ($\sigma_{\rm O-C}$\,=\,3.69\,m\,s$^{\rm -1}$; $\chi^{\rm 2}_{\rm red}$\,=\,2.89). We were 
unable to significantly improve the fit by adding extra signals. No obvious correlation between {\footnotesize CCF} bisectors and 
residuals could be detected. The origin of these abnormal residuals remains unknown, but they may be due to intrinsic stellar jitter.

The minimum mass of \object{{\footnotesize HD}\,148156}\,b is 0.85\,M$_{\rm Jup}$ and the semi-major axis of its orbit is 2.45\,AU 
(assuming $M_{*}$\,=\,1.22\,M$_{\odot}$). The fitted orbit and the fit residuals are displayed in panel~{\bf g} of Fig.~\ref{orbits}. The 
phase-folded orbit and velocities are displayed in panel {\bf e} of Fig.~\ref{phased}.

\subsection{HD\,156411}\label{hd156411}

We obtained 50 {\footnotesize HARPS} radial-velocity measurements of \object{{\footnotesize HD}\,156411} ({\footnotesize HIP}\,84787). 
These measurements were taken between {\footnotesize BJD}\,=\,2\,452\,850 (July 29, 2003) and {\footnotesize BJD}\,=\,2\,455\,081 
(September 6, 2009) and they have a mean velocity uncertainty of 1.26\,m\,s$^{\rm -1}$. A long-period signal is visible on the data 
displayed in panel~{\bf h} of Fig.~\ref{orbits}. We fitted a Keplerian orbit to these data. The resulting parameters are listed in 
Table~\ref{rvsol}. The dispersion of the residuals to the fitted orbit is too large ($\sigma_{\rm O-C}$\,=\,2.94\,m\,s$^{\rm -1}$), and 
the fit quality quite poor ($\chi^{\rm 2}_{\rm red}$\,=\,5.21).

We tried to fit models including additional signals. First, we tried to add linear and quadratic drifts. In both cases, the fit quality 
improved a bit ($\chi^{\rm 2}_{\rm red}$\,=4.80 and 4.70, respectively) but the corresponding f-test probabilities indicated that the 
achieved improvements were only marginally significant (at the 1\,$\sigma$ level). A 2-planet fit was also tried, but we were unable to 
find any convincing solution. We also verified the absence of correlation between {\footnotesize CCF} bisectors and residuals. 
\object{{\footnotesize HD}\,156411} is slightly evolved so we speculate that intrinsic stellar signals are responsible for the abnormal 
residuals.

For this paper, we finally decided to adopt our initial 1-planet orbital solution. With this solution and assuming a mass of 
1.25\,M$_{\odot}$ for the host star, we computed a minimum mass of 0.74\,M$_{\rm Jup}$ and an orbital semi-major axis of 1.88\,AU for 
\object{{\footnotesize HD}\,156411}\,b. The fitted orbit and the temporal residuals are displayed in panel~{\bf h} of Fig.~\ref{orbits}. 
The phase-folded orbit and velocities are displayed in panel {\bf f} of Fig.~\ref{phased}.

\section{Conclusions}\label{conclusion}

In this paper, we have presented our {\footnotesize HARPS} radial-velocity data for 8 low-activity solar-type stars orbited by at least 
one long-period planetary companion. \object{{\footnotesize HD}\,6718} has a 1.56\,M$_{\rm Jup}$ companion on a 2496-d low-eccentricity 
orbit ($e$\,=\,0.1). The companion orbiting \object{{\footnotesize HD}\,8535} is a 0.68\,M$_{\rm Jup}$ planet with a period of 1313 days 
and an eccentricity of 0.15. \object{{\footnotesize HD}\,28254}\,b is a 1.16\,M$_{\rm Jup}$ planet on a 1116-d highly eccentric orbit 
($e$\,=\,0.81). The companion to {\footnotesize HD}\,290327 has a minimum mass of 2.54\,M$_{\rm Jup}$. Its 2442-d orbit is nearly circular 
($e$\,=\,0.08). As for \object{{\footnotesize HD}\,28254}\,b, the 0.6\,M$_{\rm Jup}$ companion to \object{{\footnotesize HD}\,43197} is on 
a highly eccentric orbit: $e$\,=\,0.83. Its orbital period is 327\,d. \object{{\footnotesize HD}\,44219} is orbited by a 
0.58\,M$_{\rm Jup}$ planet on a 472-d 0.61 eccentricity orbit. \object{{\footnotesize HD}\,148156}\,b is a 0.85\,M$_{\rm Jup}$ planet 
orbiting its parent star on a 1027-d eccentric orbit ($e$\,=\,0.52) and the companion to \object{{\footnotesize HD}\,156411} is a 0.74 
Jupiter-mass planet with a mildly eccentric ($e$\,=\,0.22) 842-d orbit. 

We find some evidence of additional longer-period companions around \object{{\footnotesize HD}\,28254} and 
\object{{\footnotesize HD}\,44219}. In both cases, we are unable to fit full double-Keplerian orbits because of our short observing time 
span. Future observations of these two systems will probably allow us to better characterize the companions responsible for these signals.

Finally, with the addition of \object{{\footnotesize HD}\,28254}\,b and \object{{\footnotesize HD}\,43197}\,b, the population of 
exoplanets with extremely high eccentricities now amounts to 6. The other planets with $e$\,$\geq$\,0.8 are 
\object{{\footnotesize HD}\,4113} \citep[$e$\,=\,0.903, ][]{Tamuz2008}, {\footnotesize 156846} \citep[$e$\,=\,0.8472, ][]{Tamuz2008}, 
\object{{\footnotesize HD}\,20782}\,b \citep[$e$\,=\,0.92, ][]{Jones2006}, and \object{{\footnotesize HD}\,80606}\,b 
\citep[$e$\,=\,0.934, ][]{Moutou80606}. Five of these stars hosting highly eccentric planets are members of wide visual binaries (see 
\citet{Tamuz2008} for \object{{\footnotesize HD}\,4113} and \object{{\footnotesize HD}\,156846}, \citet{Desidera2007} for 
\object{{\footnotesize HD}\,20792}, and \citet{Naef2001} for \object{{\footnotesize HD}\,80606}).Only \object{{\footnotesize HD}\,43197} 
is not known to be in a multiple stellar system. This large number of eccentric-planet hosts in binaries tend to favour scenarios in which 
the extreme orbital eccentricity results from the influence of a third body via Kozai oscillations \citep{Kozai1962} or Kozai migration 
\citep{Wu2003}.

\begin{acknowledgements}\label{ackno}
We thank the Swiss National Science Foundation (SNSF) and Geneva University for their continuous support of our planet search programmes.
NCS would like to acknowledge the support from the European Research Council/European Community under the FP7 through a Starting Grant, 
as well from Funda\c{c}\~ao para a Ci\^encia e a Tecnologia (FCT), Portugal, through programme Ci\^encia\,2007, and in the form of grants 
PTDC/CTE-AST/098528/2008 and PTDC/CTE-AST/098604/2008. This research has made use of the Simbad database, operated at the 
{\footnotesize CDS} in Strasbourg, France.
\end{acknowledgements}

\bibliographystyle{aa}
\bibliography{13616}

\begin{thebibliography}{39}
\expandafter\ifx\csname natexlab\endcsname\relax\def\natexlab#1{#1}\fi

\bibitem[{{Baranne} {et~al.}(1996){Baranne}, {Queloz}, {Mayor},
  {et~al.}}]{Baranne96}
{Baranne}, A., {Queloz}, D., {Mayor}, M., {et~al.} 1996, \aaps, 119, 373

\bibitem[{{Bouchy} {et~al.}(2005){Bouchy}, {Bazot}, {Santos}, {Vauclair}, \&
  {Sosnowska}}]{Bouchy2005}
{Bouchy}, F., {Bazot}, M., {Santos}, N.~C., {Vauclair}, S., \& {Sosnowska}, D.
  2005, \aap, 440, 609

\bibitem[{{Delfosse} {et~al.}(2000){Delfosse}, {Forveille}, {S{\' e}gransan},
  {Beuzit}, {Udry}, {Perrier}, \& {Mayor}}]{Delfosse2000}
{Delfosse}, X., {Forveille}, T., {S{\' e}gransan}, D., {et~al.} 2000, \aap,
  364, 217

\bibitem[{{Desidera} \& {Barbieri}(2007)}]{Desidera2007}
{Desidera}, S. \& {Barbieri}, M. 2007, \aap, 462, 345

\bibitem[{{Dommanget} \& {Nys}(2002)}]{Dommanget2002}
{Dommanget}, J. \& {Nys}, O. 2002, IAU Commission on Double Stars, 148, 4

\bibitem[{{ESA}(1997)}]{ESA97}
{ESA}. 1997, The {\footnotesize HIPPARCOS} and {\footnotesize TYCHO} catalogue,
  ESA-SP 1200

\bibitem[{{Fernandes} \& {Santos}(2004)}]{Fernandes2004}
{Fernandes}, J. \& {Santos}, N.~C. 2004, \aap, 427, 607

\bibitem[{{Flower}(1996)}]{Flower96}
{Flower}, P.~J. 1996, \apj, 469, 355

\bibitem[{{Girardi} {et~al.}(2000){Girardi}, {Bressan}, {Bertelli}, \&
  {Chiosi}}]{Girardi2000}
{Girardi}, L., {Bressan}, A., {Bertelli}, G., \& {Chiosi}, C. 2000, \aaps, 141,
  371

\bibitem[{{H{\'e}brard} {et~al.}(2010){H{\'e}brard}, {Udry}, {Lo Curto},
  {Robichon}, {Naef}, {Ehrenreich}, {Benz}, {Bouchy}, {Lecavelier Des Etangs},
  {Lovis}, {Mayor}, {Moutou}, {Pepe}, {Queloz}, {Santos}, \&
  {S{\'e}gransan}}]{Hebrard2009}
{H{\'e}brard}, G., {Udry}, S., {Lo Curto}, G., {et~al.} 2010, \aap, 512, A46

\bibitem[{{Jenkins} {et~al.}(2008){Jenkins}, {Jones}, {Pavlenko}, {Pinfield},
  {Barnes}, \& {Lyubchik}}]{Jenkins2008}
{Jenkins}, J.~S., {Jones}, H.~R.~A., {Pavlenko}, Y., {et~al.} 2008, \aap, 485,
  571

\bibitem[{{Jones} {et~al.}(2006){Jones}, {Butler}, {Tinney}, {Marcy}, {Carter},
  {Penny}, {McCarthy}, \& {Bailey}}]{Jones2006}
{Jones}, H.~R.~A., {Butler}, R.~P., {Tinney}, C.~G., {et~al.} 2006, \mnras,
  369, 249

\bibitem[{{Kozai}(1962)}]{Kozai1962}
{Kozai}, Y. 1962, \aj, 67, 591

\bibitem[{{Lo Curto} {et~al.}(2010){Lo Curto}, {Mayor}, {Benz}, {Bouchy},
  {Lovis}, {Moutou}, {Naef}, {Pepe}, {Queloz}, {Santos}, {Segransan}, \&
  {Udry}}]{Locurto2009}
{Lo Curto}, G., {Mayor}, M., {Benz}, W., {et~al.} 2010, \aap, 512, A48

\bibitem[{{Lo Curto} {et~al.}(2006){Lo Curto}, {Mayor}, {Clausen}, {Benz},
  {Bouchy}, {Lovis}, {Moutou}, {Naef}, {Pepe}, {Queloz}, {Santos}, {Sivan},
  {Udry}, {Bonfils}, {Delfosse}, {Mordasini}, {Fouqu{\'e}}, {Olsen}, \&
  {Pritchard}}]{Locurto2006}
{Lo Curto}, G., {Mayor}, M., {Clausen}, J.~V., {et~al.} 2006, \aap, 451, 345

\bibitem[{{Mayor} {et~al.}(2009){Mayor}, {Bonfils}, {Forveille}, {Delfosse},
  {Udry}, {Bertaux}, {Beust}, {Bouchy}, {Lovis}, {Pepe}, {Perrier}, {Queloz},
  \& {Santos}}]{Mayor2009}
{Mayor}, M., {Bonfils}, X., {Forveille}, T., {et~al.} 2009, \aap, 507, 487

\bibitem[{{Mayor} {et~al.}(2003){Mayor}, {Pepe}, {Queloz}, {Bouchy},
  {Rupprecht}, {Lo Curto}, {Avila}, {Benz}, {Bertaux}, {Bonfils}, {dall},
  {Dekker}, {Delabre}, {Eckert}, {Fleury}, {Gilliotte}, {Gojak}, {Guzman},
  {Kohler}, {Lizon}, {Longinotti}, {Lovis}, {Megevand}, {Pasquini}, {Reyes},
  {Sivan}, {Sosnowska}, {Soto}, {Udry}, {van Kesteren}, {Weber}, \&
  {Weilenmann}}]{Mayor2003}
{Mayor}, M., {Pepe}, F., {Queloz}, D., {et~al.} 2003, The Messenger, 114, 20

\bibitem[{{Mordasini} {et~al.}(2009){Mordasini}, {Alibert}, {Benz}, \&
  {Naef}}]{Mordasini2009}
{Mordasini}, C., {Alibert}, Y., {Benz}, W., \& {Naef}, D. 2009, \aap, 501, 1161

\bibitem[{{Mordasini} {et~al.}(2010){Mordasini}, {Mayor}, {Udry}, {Lovis},
  {S\'egransan}, {Benz}, {Bertaux}, {Bouchy}, {Lo Curto}, {Moutou}, {Naef},
  {Pepe}, {Queloz}, \& {Santos}}]{MordasiniHARPS}
{Mordasini}, C., {Mayor}, M., {Udry}, S., {et~al.} 2010, \aap, Submitted

\bibitem[{{Moutou} {et~al.}(2009{\natexlab{a}}){Moutou}, {H{\'e}brard},
  {Bouchy}, {Eggenberger}, {Boisse}, {Bonfils}, {Gravallon}, {Ehrenreich},
  {Forveille}, {Delfosse}, {Desort}, {Lagrange}, {Lovis}, {Mayor}, {Pepe},
  {Perrier}, {Pont}, {Queloz}, {Santos}, {S{\'e}gransan}, {Udry}, \&
  {Vidal-Madjar}}]{Moutou80606}
{Moutou}, C., {H{\'e}brard}, G., {Bouchy}, F., {et~al.} 2009{\natexlab{a}},
  \aap, 498, L5

\bibitem[{{Moutou} {et~al.}(2005){Moutou}, {Mayor}, {Bouchy}, {Lovis}, {Pepe},
  {Queloz}, {Santos}, {Udry}, {Benz}, {Lo Curto}, {Naef}, {S{\'e}gransan}, \&
  {Sivan}}]{Moutou2005}
{Moutou}, C., {Mayor}, M., {Bouchy}, F., {et~al.} 2005, \aap, 439, 367

\bibitem[{{Moutou} {et~al.}(2009{\natexlab{b}}){Moutou}, {Mayor}, {Lo Curto},
  {Udry}, {Bouchy}, {Benz}, {Lovis}, {Naef}, {Pepe}, {Queloz}, \&
  {Santos}}]{Moutou2009}
{Moutou}, C., {Mayor}, M., {Lo Curto}, G., {et~al.} 2009{\natexlab{b}}, \aap,
  496, 513

\bibitem[{{Moutou} {et~al.}(2010){Moutou}, {Mayor}, {Lo Curto}, {Udry},
  {Bouchy}, {Benz}, {Lovis}, {Naef}, {Pepe}, {Queloz}, {Santos}, \&
  {Sousa}}]{Moutou2010}
{Moutou}, C., {Mayor}, M., {Lo Curto}, G., {et~al.} 2010, A\&A, Submitted

\bibitem[{{Naef} {et~al.}(2001){Naef}, {Latham}, {Mayor}, {Mazeh}, {Beuzit},
  {Drukier}, {Perrier-Bellet}, {Queloz}, {Sivan}, {Torres}, {Udry}, \&
  {Zucker}}]{Naef2001}
{Naef}, D., {Latham}, D.~W., {Mayor}, M., {et~al.} 2001, \aap, 375, L27

\bibitem[{{Naef} {et~al.}(2007){Naef}, {Mayor}, {Benz}, {Bouchy}, {Lo Curto},
  {Lovis}, {Moutou}, {Pepe}, {Queloz}, {Santos}, \& {Udry}}]{Naef2007}
{Naef}, D., {Mayor}, M., {Benz}, W., {et~al.} 2007, \aap, 470, 721

\bibitem[{{Pace} \& {Pasquini}(2004)}]{Pace2004}
{Pace}, G. \& {Pasquini}, L. 2004, \aap, 426, 1021

\bibitem[{{Pepe} {et~al.}(2004{\natexlab{a}}){Pepe}, {Mayor}, {Queloz}, {Benz},
  {Bonfils}, {Bouchy}, {Curto}, {Lovis}, {M{\'e}gevand}, {Moutou}, {Naef},
  {Rupprecht}, {Santos}, {Sivan}, {Sosnowska}, \& {Udry}}]{Pepe2004}
{Pepe}, F., {Mayor}, M., {Queloz}, D., {et~al.} 2004{\natexlab{a}}, \aap, 423,
  385

\bibitem[{{Pepe} {et~al.}(2004{\natexlab{b}}){Pepe}, {Mayor}, {Queloz}, \&
  {Udry}}]{PepeIAU}
{Pepe}, F., {Mayor}, M., {Queloz}, D., \& {Udry}, S. 2004{\natexlab{b}}, in IAU
  Symposium, Vol. 202, Planetary Systems in the Universe, ed. {A.~Penny}, 103

\bibitem[{{Robinson} {et~al.}(2007){Robinson}, {Ammons}, {Kretke}, {Strader},
  {Wertheimer}, {Fischer}, \& {Laughlin}}]{Robinson2007}
{Robinson}, S.~E., {Ammons}, S.~M., {Kretke}, K.~A., {et~al.} 2007, \apjs, 169,
  430

\bibitem[{{Santos} {et~al.}(2004{\natexlab{a}}){Santos}, {Bouchy}, {Mayor},
  {Pepe}, {Queloz}, {Udry}, {Lovis}, {Bazot}, {Benz}, {Bertaux}, {Lo Curto},
  {Delfosse}, {Mordasini}, {Naef}, {Sivan}, \& {Vauclair}}]{Santosmuarae}
{Santos}, N.~C., {Bouchy}, F., {Mayor}, M., {et~al.} 2004{\natexlab{a}}, \aap,
  426, L19

\bibitem[{{Santos} {et~al.}(2004{\natexlab{b}}){Santos}, {Israelian}, \&
  {Mayor}}]{Santos2004}
{Santos}, N.~C., {Israelian}, G., \& {Mayor}, M. 2004{\natexlab{b}}, \aap, 415,
  1153

\bibitem[{{Santos} {et~al.}(2000){Santos}, {Mayor}, {Naef}, {Pepe}, {Queloz},
  {Udry}, \& {Blecha}}]{Santos2000}
{Santos}, N.~C., {Mayor}, M., {Naef}, D., {et~al.} 2000, \aap, 361, 265

\bibitem[{{Santos} {et~al.}(2002){Santos}, {Mayor}, {Naef}, {Pepe}, {Queloz},
  {Udry}, {Burnet}, {Clausen}, {Helt}, {Olsen}, \& {Pritchard}}]{Santos2002}
{Santos}, N.~C., {Mayor}, M., {Naef}, D., {et~al.} 2002, \aap, 392, 215

\bibitem[{{S\'egransan} {et~al.}(2010){S\'egransan}, {Mayor}, {Udry}, {Lovis},
  {Benz}, {Bouchy}, {Lo Curto}, {Mordasini}, {Moutou}, {Naef}, {Pepe},
  {Queloz}, \& {Santos}}]{Segransan2009}
{S\'egransan}, D., {Mayor}, M., {Udry}, S., {et~al.} 2010, \aap, Submitted

\bibitem[{{Sousa} {et~al.}(2007){Sousa}, {Santos}, {Israelian}, {Mayor}, \&
  {Monteiro}}]{Sousa2007}
{Sousa}, S.~G., {Santos}, N.~C., {Israelian}, G., {Mayor}, M., \& {Monteiro},
  M.~J.~P.~F.~G. 2007, \aap, 469, 783

\bibitem[{{Tamuz} {et~al.}(2008){Tamuz}, {S{\'e}gransan}, {Udry}, {Mayor},
  {Eggenberger}, {Naef}, {Pepe}, {Queloz}, {Santos}, {Demory}, {Figuera},
  {Marmier}, \& {Montagnier}}]{Tamuz2008}
{Tamuz}, O., {S{\'e}gransan}, D., {Udry}, S., {et~al.} 2008, \aap, 480, L33

\bibitem[{{Udry} {et~al.}(2000){Udry}, {Mayor}, {Queloz}, {Naef}, \&
  {Santos}}]{Udrysample}
{Udry}, S., {Mayor}, M., {Queloz}, D., {Naef}, D., \& {Santos}, N. 2000, in
  From Extrasolar Planets to Cosmology: The VLT Opening Symposium, ed.
  {J.~Bergeron \& A.~Renzini}, 571

\bibitem[{{van Leeuwen}(2007)}]{vanleeuwen2007}
{van Leeuwen}, F. 2007, \aap, 474, 653

\bibitem[{{Wu} \& {Murray}(2003)}]{Wu2003}
{Wu}, Y. \& {Murray}, N. 2003, \apj, 589, 605

\end{thebibliography}
\end{document}